\documentclass[12pt]{article}
\usepackage{a4wide,epsfig,psfrag,amsmath,amssymb,cite,scalefnt}
\usepackage{color}

\graphicspath{ {figs/} }

\newcommand{\ep}{\epsilon}

\newcommand{\nl}{n_l}
\newcommand{\logx}{l_x}
\newcommand{\cR}{C_F}
\newcommand{\cA}{C_A}
\newcommand{\nh}{n_h}
\newcommand{\tf}{T_F}
\newcommand{\zthree}{\zeta(3)}
\newcommand{\logtwo}{l_2}
\newcommand{\afour}{a_4}

\allowdisplaybreaks

\begin{document}

\title{\vskip-3cm{\baselineskip14pt
    \begin{flushleft}
     \normalsize TTP18-006
    \end{flushleft}} \vskip1.5cm 
  Three-loop massive form factors: complete light-fermion corrections for the
  vector current
  }

\author{
  Roman N. Lee$^{a}$,
  Alexander V. Smirnov$^{b}$,
  \\
  Vladimir A. Smirnov$^{c,d}$,
  Matthias Steinhauser$^{d}$,
  \\[1em]
  {\small\it (a) Budker Institute of Nuclear Physics,}\\
  {\small\it  630090 Novosibirsk, Russia}
  \\
  {\small\it (b) Research Computing Center, Moscow State University}\\
  {\small\it 119991, Moscow, Russia}
  \\  
  {\small\it (c) Skobeltsyn Institute of Nuclear Physics of Moscow State University}\\
  {\small\it 119991, Moscow, Russia}
  \\
  {\small\it (d) Institut f{\"u}r Theoretische Teilchenphysik,
    Karlsruhe Institute of Technology (KIT)}\\
  {\small\it 76128 Karlsruhe, Germany}  
}
  
\date{}

\maketitle

\thispagestyle{empty}

\begin{abstract}

  We compute the three-loop QCD corrections to the massive
  quark-anti-quark-photon form factors $F_1$ and $F_2$ involving a closed loop
  of massless fermions.  This subset is gauge invariant and contains both
  planar and non-planar contributions.  We perform the reduction using {\tt FIRE}
  and compute the master integrals with the help of differential equations.
  Our analytic results can be expressed in terms of Goncharov polylogarithms.
  We provide analytic results for all master integrals which are
  not present in the large-$N_c$ calculation considered in
  Refs.~\cite{Henn:2016kjz,Henn:2016tyf}.

%

\end{abstract}

\thispagestyle{empty}


\newpage


\section{Introduction}

In the absence of striking experimental signals which hint to physics beyond
the Standard Model it is of utmost importance to increase the precision of the
theoretical predictions. A subsequent detailed comparison to precise
measurements will help to uncover deviations and will provide hints
for the construction of beyond-the-Standard-Model theories.

Quark and gluon 
form factors play a special role in the context of precision calculations. On
the one hand they are sufficiently simple which allows to compute them to high
order in perturbation theory. On the other hand they enter as building blocks
into a variety of physical cross sections and decay rates, most prominently
into Higgs boson production and decay, the Drell Yan production of leptons and
the production of massive quarks. Form factors also constitute an ideal
playground to study the infrared properties of a quantum field theory, in
particular of QCD.  As far as massless form factors are concerned the
state-of-the-art is four loops where different groups have contributed to
partial
results~\cite{Henn:2016men,Lee:2016ixa,Lee:2017mip,vonManteuffel:2016xki,Boels:2017skl,Boels:2017fng}.
Massive quark form factors are known to two loops~\cite{Bernreuther:2004ih}
including ${\cal O}(\epsilon)$~\cite{Gluza:2009yy,Henn:2016tyf} and ${\cal
  O}(\epsilon^2)$ terms~\cite{Ahmed:2017gyt,Ablinger:2017hst}.  Three-loop
corrections in the large-$N_c$ limit for the vector current form factor have
been computed in Ref.~\cite{Henn:2016tyf}.  In this paper we extend these
considerations and compute the complete contributions (i.e. all colour
factors) from the diagrams involving a closed massless quark loop. This
well-defined and gauge invariant subset contains for the first time non-planar
contributions which we study in detail.  Furthermore, new planar master
integrals have to be evaluated which are not present in the large-$N_c$
result. As a by-product of our calculation we obtain the two-loop form factor
including order $\epsilon^2$ terms.
We do not consider singlet diagrams where the external photon couples to
a closed massless quark loop which is connected via gluons to the final-state massive
quarks. Such diagrams form again a separate gauge invariant subset which
requires the computation of different integral families.
Let us mention that all-order corrections to the massive form factor
in the large-$\beta_0$ limit have been considered in Ref.~\cite{Grozin:2017aty}.

The remainder of the paper is structured as follows: In the next
Section we introduce the notation and discuss the ultraviolet and
infrared divergences. One- and two-loop results are presented in
Section~\ref{sec:12loops}. The three-loop calculation is described in
Section~\ref{sec::3lres}, in particular the calculation of the master
integrals.  Section~\ref{sec::num} contains a discussion of the
three-loop form factor.  We provide both numerical results and
analytic expressions in various kinematical limits.  In
Section~\ref{sec::sum} we summarize our results and comment on the
perspective for the full result.


\section{Notation, renormalization and infrared structure}

Let us define the form factors we are going to consider.
Starting point is the photon-quark-anti-quark vertex which we introduce as
\begin{eqnarray}
  V^{\mu,ij}(q_1,q_2) &=& \delta^{ij} \, \bar{u}(q_1) \Gamma^\mu(q_1,q_2) v(q_2)
  \,,
  \label{eq::Vmu}
\end{eqnarray}
where $i$ and $j$ are (fundamental) colour indices and
$\bar{u}(q_1)$ and $v(q_2)$ are the spinors of the quark and anti-quark,
respectively, with incoming momentum $q_1$ and outgoing momentum $q_2$.
The external quarks are on-shell, i.e., we have $q_1^2=q_2^2=m^2$.
The form factors are defined as prefactors of the Lorentz decomposition of the
vertex function $\Gamma^\mu(q_1,q_2)$ which is introduced as
\begin{eqnarray}
  \Gamma^\mu(q_1,q_2) &=& Q_q
  \left[F_1(q^2)\gamma^\mu - \frac{i}{2m}F_2(q^2) \sigma^{\mu\nu}q_\nu\right]
  \,,
  \label{eq::Gamma}
\end{eqnarray}
with $q=q_1-q_2$ being the outgoing momentum of the photon and
$\sigma^{\mu\nu} = i[\gamma^\mu,\gamma^\nu]/2$.  $Q_q$ is the charge of the
considered quark. For on-shell renormalized form factors we have
$F_1(0)=1$ and $F_2(0)=(g-2)/2$ where $g$ is the gyromagnetic ratio of the
quark (or lepton in the case of QED).
For later convenience we define the perturbative expansion of $F_1$
and $F_2$ as
\begin{eqnarray}
  F_i = \sum_{n\ge0} F_i^{(n)}
  \left(\frac{\alpha_s(\mu)}{4\pi}\right)^n
  \,,
\end{eqnarray}
with $F_1^{(0)}=1$ and $F_2^{(0)}=0$.

To obtain the renormalized form factors we use the $\overline{\rm MS}$ scheme
for the strong coupling constant and the on-shell scheme for the heavy quark
mass and wave function of the external quarks. In all cases the counterterm
contributions are simply obtained by re-scaling the bare parameters with the
corresponding renormalization constants, $Z_{\alpha_s}$, $Z_m^{\rm OS}$ and
$Z_2^{\rm OS}$. The latter is needed to three loops whereas
two-loop corrections for $Z_{\alpha_s}$ and $Z_m^{\rm
  OS}$ are sufficient to obtain renormalized three-loop results for
$F_1$ and $F_2$. Note that higher order $\epsilon$ coefficients are needed
for the on-shell renormalization constants since the one- and two-loop
form factors develop $1/\epsilon$ and $1/\epsilon^2$ poles, respectively.

After renormalization of the ultraviolet divergences the form factors still
contain infrared poles which are connected to the cusp anomalous
dimension, $\Gamma_{\rm cusp}$~\cite{Polyakov:1980ca,Brandt:1981kf,Korchemsky:1987wg}.
We adapt the notation from Ref.~\cite{Henn:2016tyf} and write
\begin{eqnarray}
  F = Z F^{f}\,,
\end{eqnarray}
where the factor $Z$, which is defined in the $\overline{\rm MS}$ scheme 
  and thus only contains poles in $\epsilon$,
absorbs the infrared divergences and $F^{f}$ is finite.  The coefficients of
the poles of $Z$ are determined by the QCD beta function and $\Gamma_{\rm
  cusp}$. In fact, the $1/\epsilon^1$ pole of the $\alpha_s^n$ term of
$Z$ is proportional to the $n$-loop correction to $\Gamma_{\rm
  cusp}$ (see, e.g., Ref.~\cite{Henn:2016tyf}.\footnote{Note that there is a
  typo in the second equation of Eq.~(12) in Ref.~\cite{Henn:2016tyf}: a
  factor ``2'' is missing in front of $\Gamma^{(1)}_{\rm cusp}$ inside the
  round brackets. The corrected equation reads $z_{2,2}=
  \Gamma^{(1)}_{\rm cusp} ( \beta_0 + 2 \Gamma^{(1)}_{\rm cusp})/16$.})

A dedicated calculation of $\Gamma_{\rm cusp}$ to three loops has been
performed in
Refs.~\cite{Polyakov:1980ca,Korchemsky:1987wg,Grozin:2014hna,Grozin:2015kna}.
An independent cross check of the large-$N_c$ result has been provided in
Ref.~\cite{Henn:2016tyf}.  In this work we reproduce all $n_l$ terms at
three-loop order by extracting $\Gamma_{\rm cusp}$ from the pole part of
the form factors.

\begin{figure}[t] 
  \begin{center}
    \includegraphics[width=\textwidth]{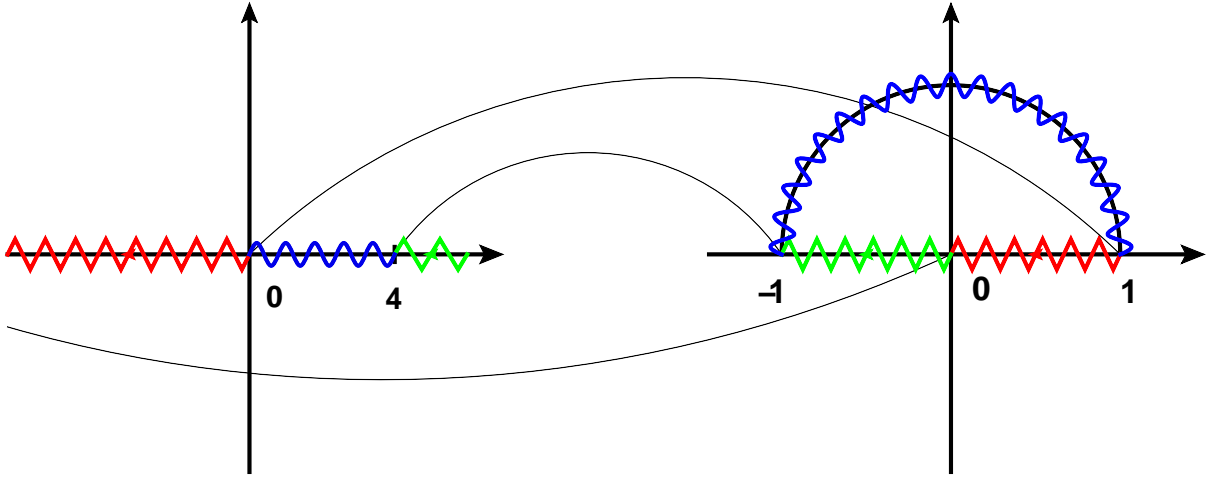}
    \caption{\label{fig::trans_x_q}Illustration of the variable transformation
      between $q^2/m^2$ and $x$ as given in Eq.~(\ref{eq::trans_x_q}).
      The left graph represents the $q^2/m^2$ plane and on right the
      complex $x$ plane is shown. The straight lines indicate the mapping for special
      values of $q^2/m^2$ and $x$.}
  \end{center}
\end{figure}

For the practical computation of the master integrals, for the
discussion of the various kinematic limits and also for the numerical
evaluation it is convenient to introduce the dimensionless
variable
\begin{eqnarray}
  \frac{q^2}{m^2} &=& - \frac{(1-x)^2}{x}
  \,,
  \label{eq::trans_x_q}
\end{eqnarray}
which maps the complex $q^2/m^2$ plane into the unit circle.  The
low-energy ($q^2\to0$), high-energy ($q^2\to\infty$) and threshold
($q^2\to4m^2$) limits correspond to $x\to 1$, $x\to 0$ and $x\to -1$,
respectively.  Furthermore, as can be seen in
Fig.~\ref{fig::trans_x_q}, the interval $q^2<0$ is mapped to $x\in(0,1)$
and $q^2\in[0,4m^2]$ to the upper semi-circle. For these values of
$x$ the form factors have to be real-valued since the corresponding
Feynman diagrams do not have cuts. This is different for the
region $q^2>4m^2$, which corresponds to $x\in(-1,0)$, where the form
factors are complex-valued.

For the threshold limit it is also convenient to introduce the velocity of the
produced quarks
\begin{eqnarray}
  \beta &=& \sqrt{1 - \frac{4 m^2}{s}}
  \,,
\end{eqnarray}
which is related to $x$ via
\begin{eqnarray}
  x &=& \frac{2\beta}{1+\beta} - 1
  \,.
\end{eqnarray}
  
\section{\label{sec:12loops}One- and two-loop form factors}

\begin{figure}[t] 
  \begin{center}
    \includegraphics[width=0.2\textwidth]{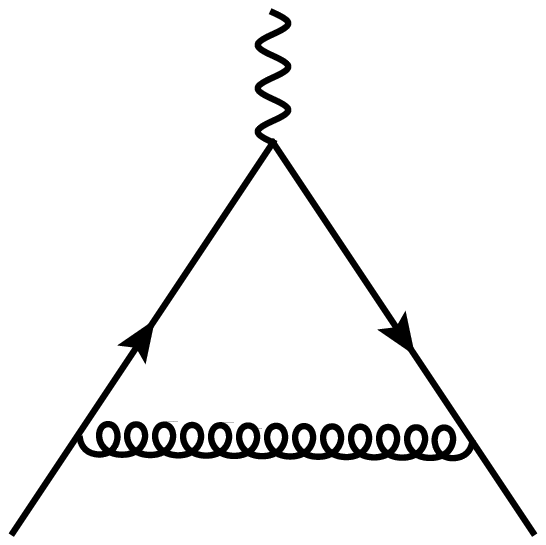} \hfill
    \includegraphics[width=0.2\textwidth]{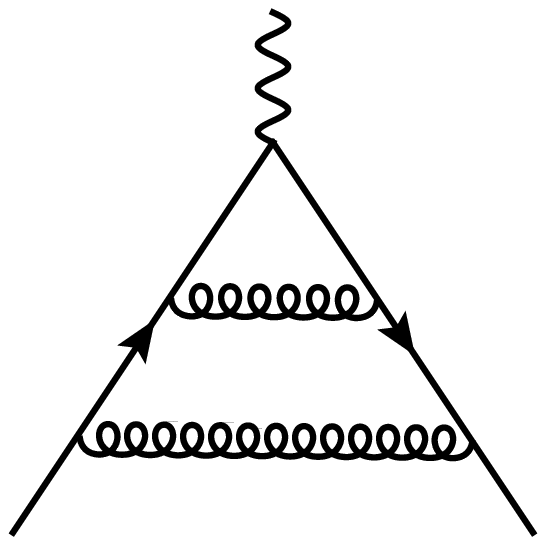}\hfill
    \includegraphics[width=0.2\textwidth]{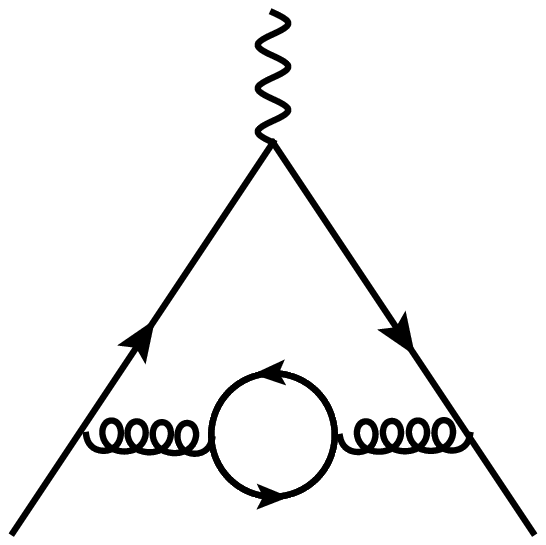}\hfill
    \includegraphics[width=0.2\textwidth]{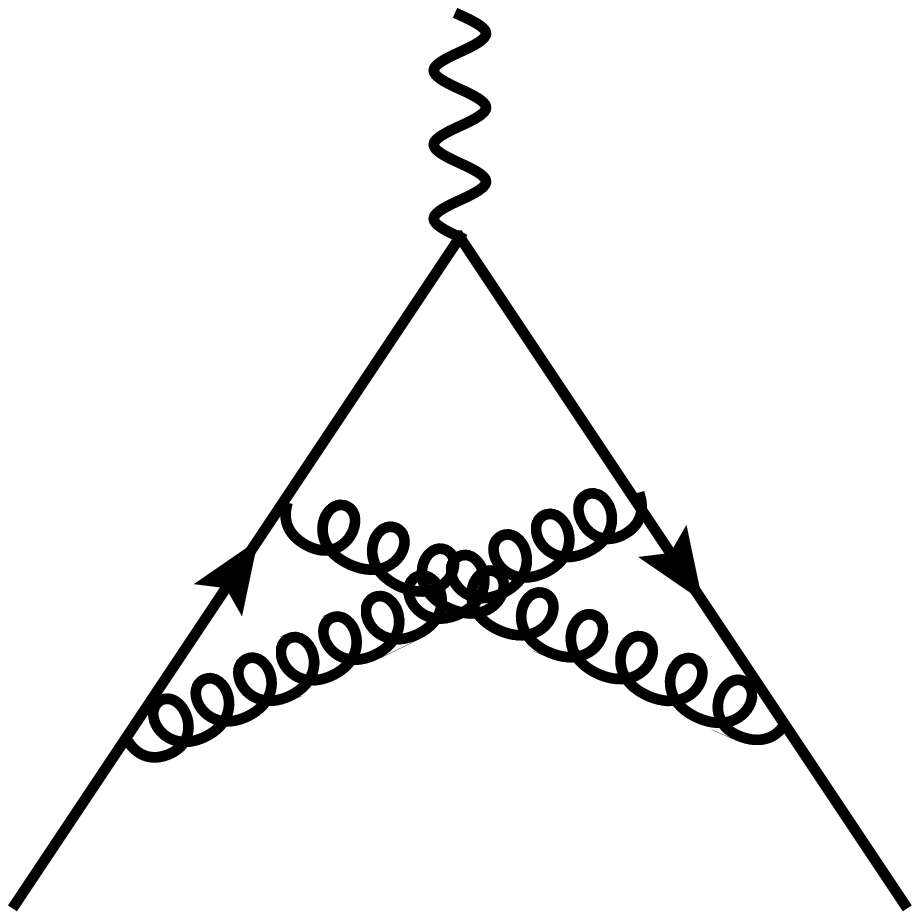}
    \caption{\label{fig::diags12}Sample diagrams contributing to $F_1$ and $F_2$
    at one and two loops. Solid, curly and wavy lines represent
    quarks, gluons and photons, respectively.}
  \end{center}
\end{figure}

\begin{figure}[t] 
  \begin{center}
    \begin{tabular}{cccc}
      \includegraphics[width=0.13\textwidth]{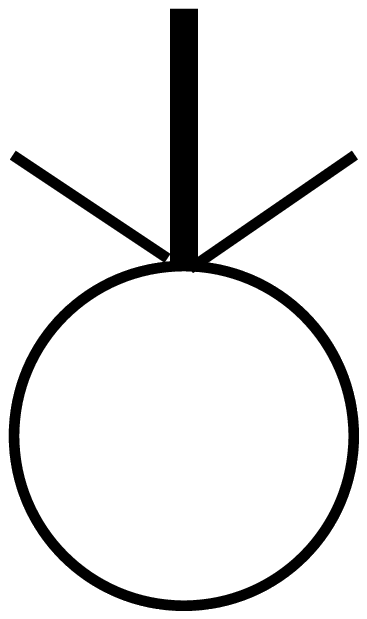}\mbox{}\hspace*{1em}\mbox{}
      & \mbox{}\hspace*{1em}\mbox{}
      \includegraphics[width=0.13\textwidth]{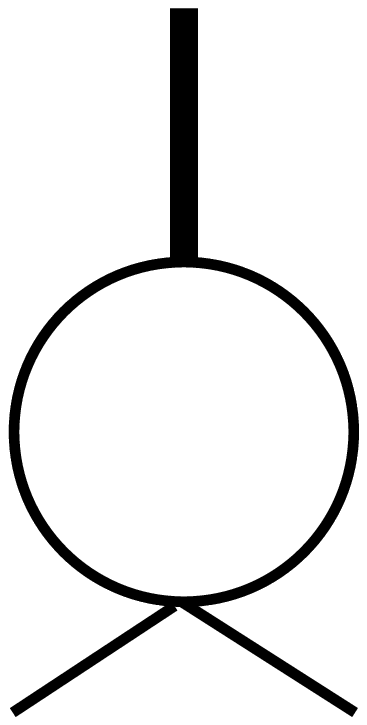}\mbox{}\hspace*{1em}\mbox{}
      & \mbox{}\hspace*{1em}\mbox{}
      \includegraphics[width=0.13\textwidth]{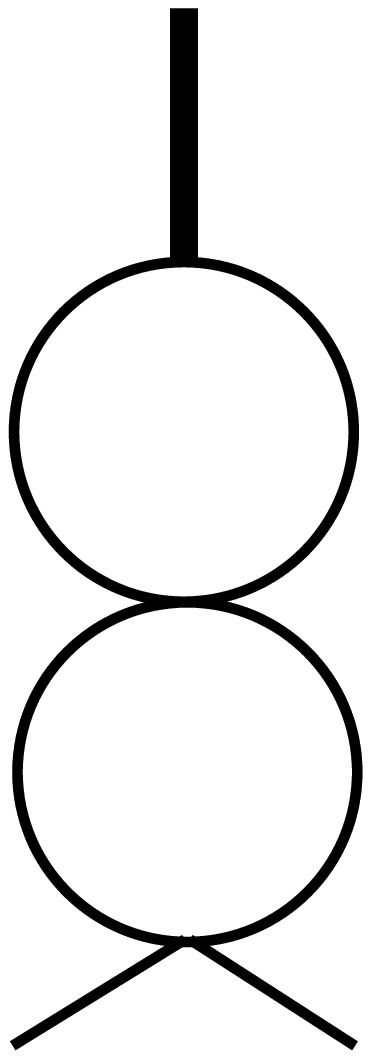}\mbox{}\hspace*{1em}\mbox{}
      & \mbox{}\hspace*{1em}\mbox{}
      \includegraphics[width=0.3\textwidth]{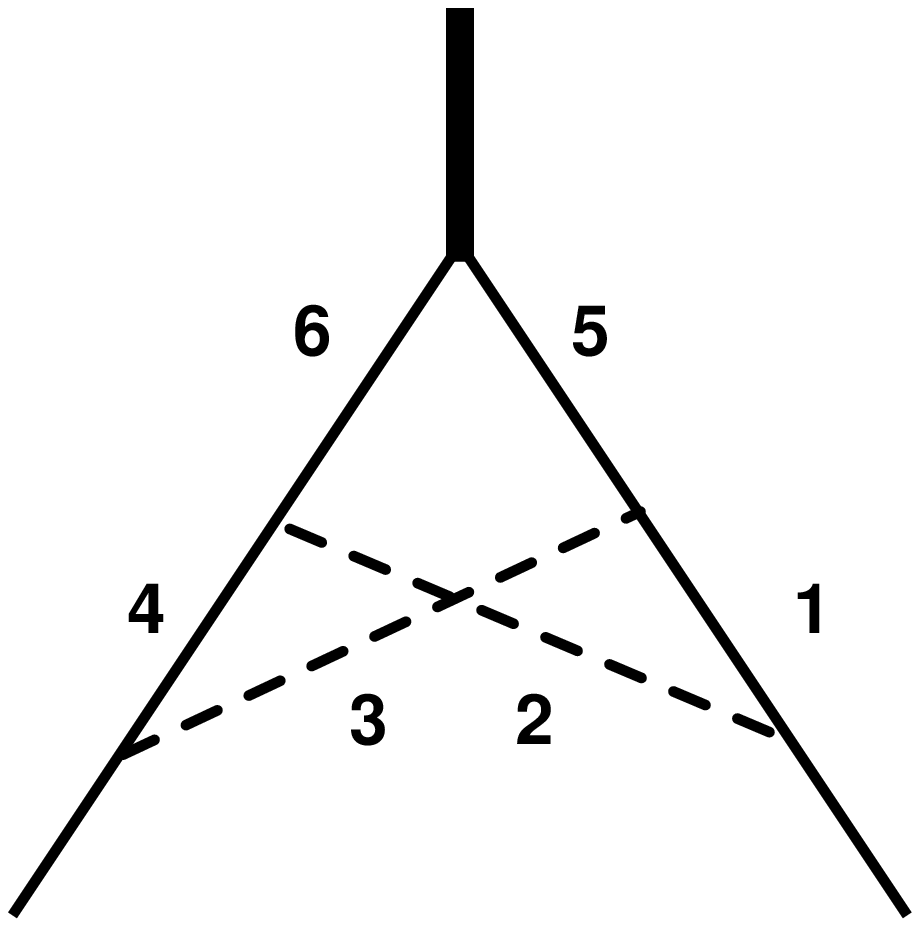}
      \\
      (a) & (b) & (c) & (d)
    \end{tabular}
    \\
    \caption{\label{fig::masters12}The two one-loop master integrals
      are shown in (a) and (b). One of the 17 master two-loop integrals
      is shown in (c) and the remaining 16 master integrals
      are obtained from (d) as described in the text.
      Solid and dashed internal lines correspond to massive and massless
      scalar propagators. Thin external lines are on the mass shell and thick
      external lines carry the (off-shell) momentum $q$.}
  \end{center}
\end{figure}

Let us in the following briefly outline the main steps of the two-loop
calculation. Sample Feynman diagrams contributing to $F_1$ and $F_2$
can be found in Fig.~\ref{fig::diags12}.  After generating the amplitudes we
find it convenient to define one integral family at one and four integral
families at two loops. We use FIRE~\cite{Smirnov:2014hma} in combination with
{\tt LiteRed}~\cite{Lee:2012cn,Lee:2013mka} for the reduction to master
integrals within each family. After minimization we arrive at two and 17
master integrals at one- and two-loop order, respectively. For convenience we
show the two one-loop and one two-loop master integrals explicitly in
Fig.~\ref{fig::masters12}(a), (b) and (c).  
The remaining 16 two-loop integrals are obtained
from ~\ref{fig::masters12}(d) by reducing lines or adding dots according to
\begin{eqnarray}
  &&G(0, 0, 0, 1, 0, 1, 0),\,\,
  G(0, 0, 0, 1, 1, 1, 0),\,\,
  G(0, 1, 0, 1, 1, 0, 0),\,\,
  \nonumber\\
  &&G(0, 1, 0, 1, 2, 0, 0),\,\, 
  G(0, 1, 1, 0, 0, 1, 0),\,\,
  G(0, 1, 1, 0, 1, 1, 0),\,\,
  \nonumber\\
  &&G(0, 1, 1, 1, 1, 0, 0),\,\,
  G(0, 1, 1, 1, 1, 1, 0),\,\, 
  G(0, 1, 1, 1, 1, 2, 0),\,\,
  \nonumber\\
  &&G(1, 0, 0, 1, 0, 1, 0),\,\, 
  G(1, 0, 0, 1, 1, 1, 0),\,\,
  G(1, 0, 0, 1, 1, 2, 0),\,\,
  \nonumber\\
  &&G(1, 0, 1, 1, 0, 1, 0),\,\,
  G(1, 0, 1, 1, 0, 2, 0),\,\, 
  G(1, 1, 1, 1, 1, 1, 0),\,\,
  \nonumber\\
  &&G(1, 1, 1, 1, 1, 2, 0)
  \,.
\end{eqnarray}
In the large-$N_c$ limit only ten master integrals are needed at two loops.

We evaluate all one- and two-loop master integrals analytically and expand in
$\epsilon$ up to the order needed for the $\epsilon^4$ and $\epsilon^2$ terms
of the one- and two-loop form factors, respectively.
Our results are expressed in terms of Goncharov polylogarithms
(GPLs)~\cite{Goncharov:1998kja} with letters $-1, 0$ and $+1$.  
  We compared the ultraviolet-renormalized
two-loop form factors to Ref.~\cite{Gluza:2009yy} and find agreement including
order $\epsilon^1$ up to the discrepancy in $F_1$ already discussed in
Section~4.4 of Ref.~\cite{Henn:2016tyf}, see also Ref.~\cite{Ablinger:2017hst}
where agreement with our result is found.  The order $\epsilon^2$ terms of
$F_1$ and $F_2$ have recently been published in Ref.~\cite{Ablinger:2017hst};
our results agree with theirs.  Note that the large-$N_c$ limit of our result
for $F_1$ has already been published in Ref.~\cite{Ahmed:2017gyt}. In this
paper the $\epsilon^2$ terms have been used to derive higher-loop corrections
with the help of renormalization group techniques. Apart from that, the
$\epsilon^2$ terms also enter a future four-loop calculation of the
massive form factors.


\section{\label{sec::3lres}Three-loop form factor}

\begin{figure}[t] 
  \begin{center}
    \begin{tabular}{ccccc}
      \includegraphics[width=0.17\textwidth]{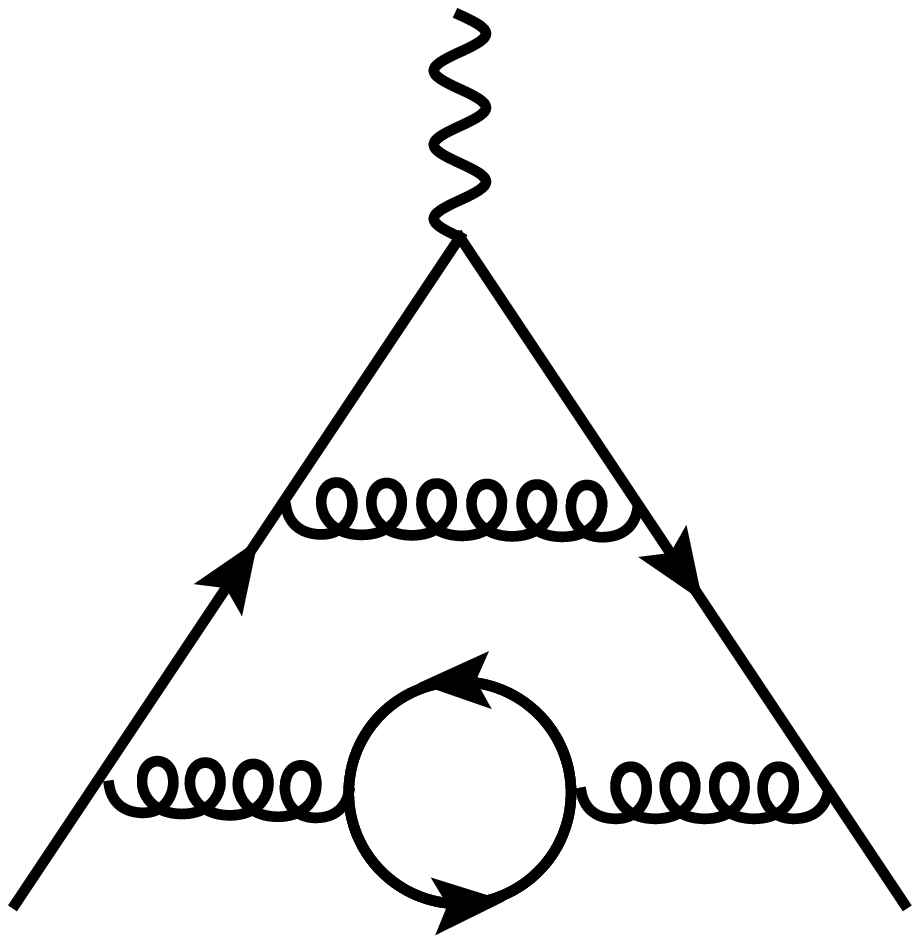} &
      \includegraphics[width=0.17\textwidth]{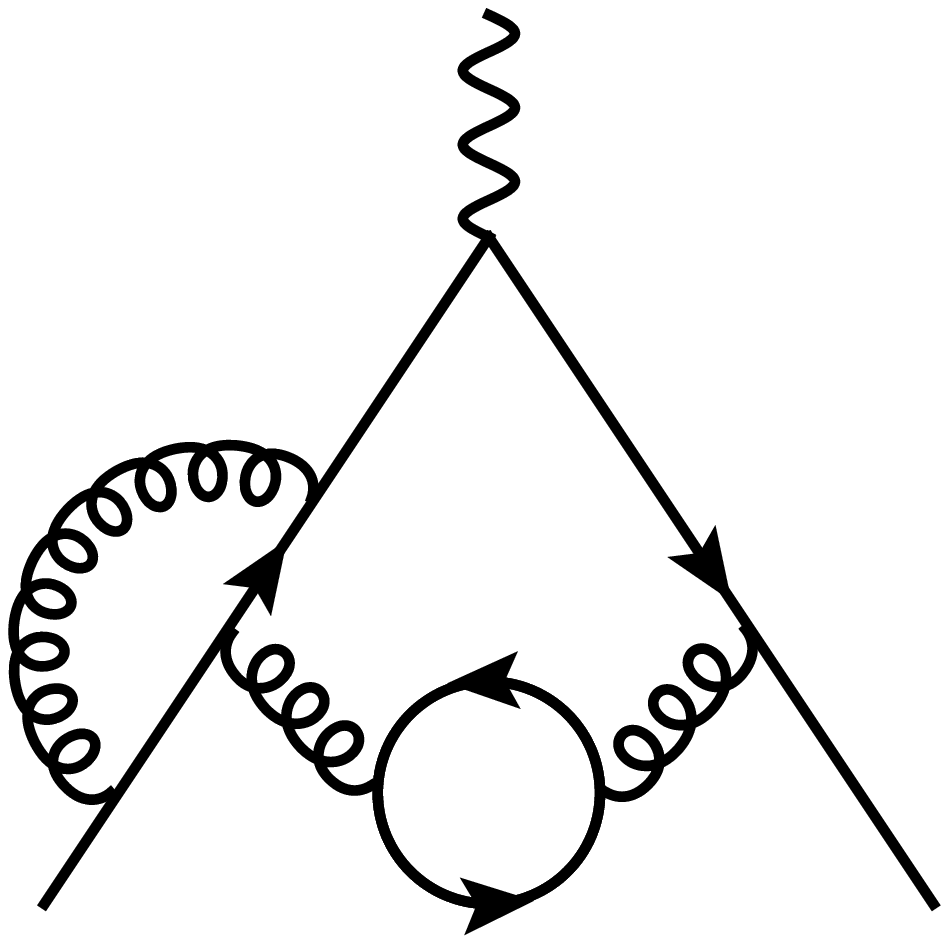} &
      \includegraphics[width=0.17\textwidth]{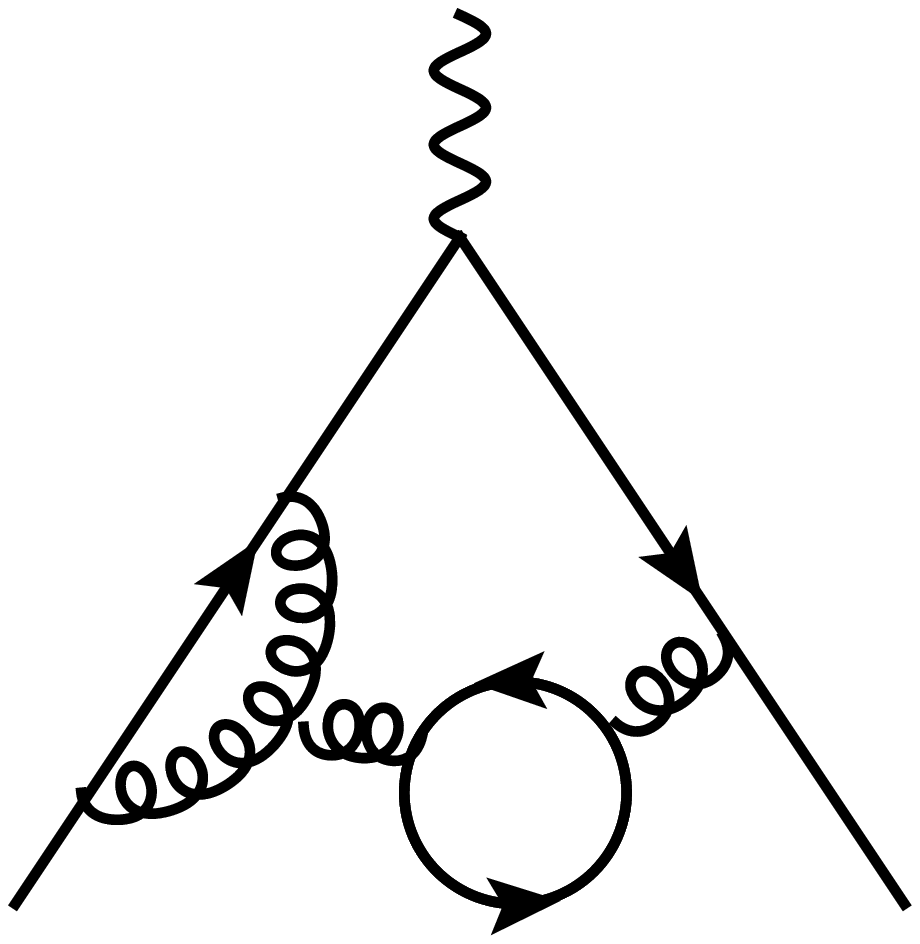} &
      \includegraphics[width=0.17\textwidth]{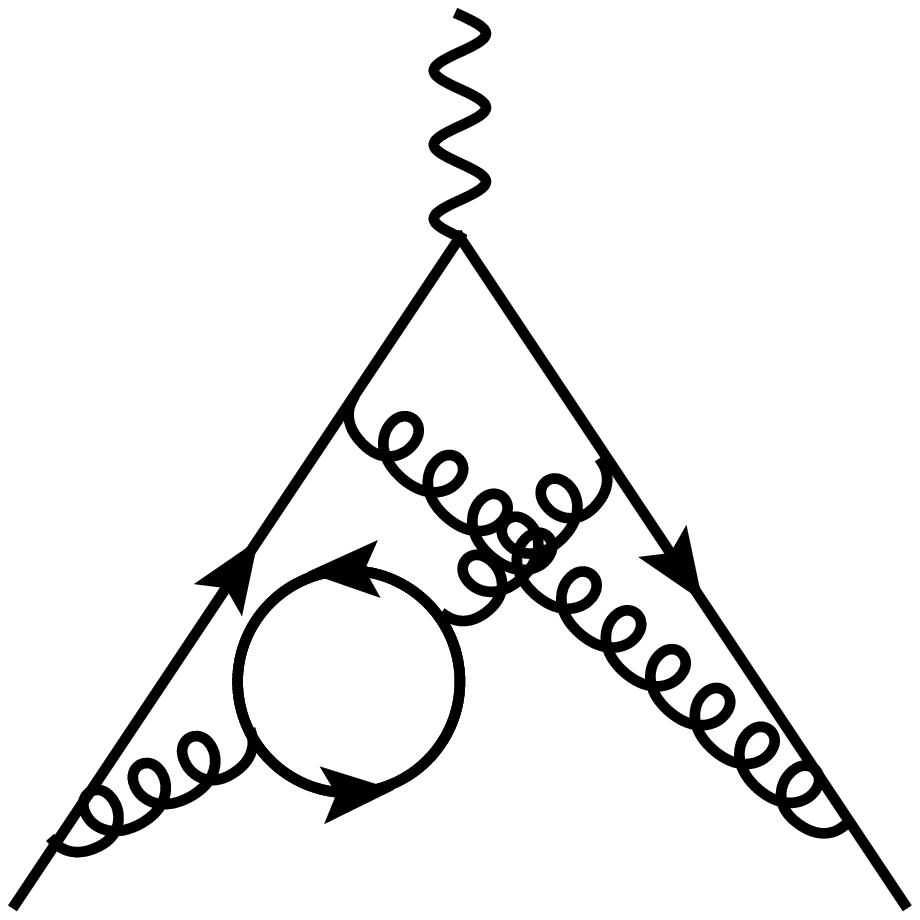} &
      \includegraphics[width=0.17\textwidth]{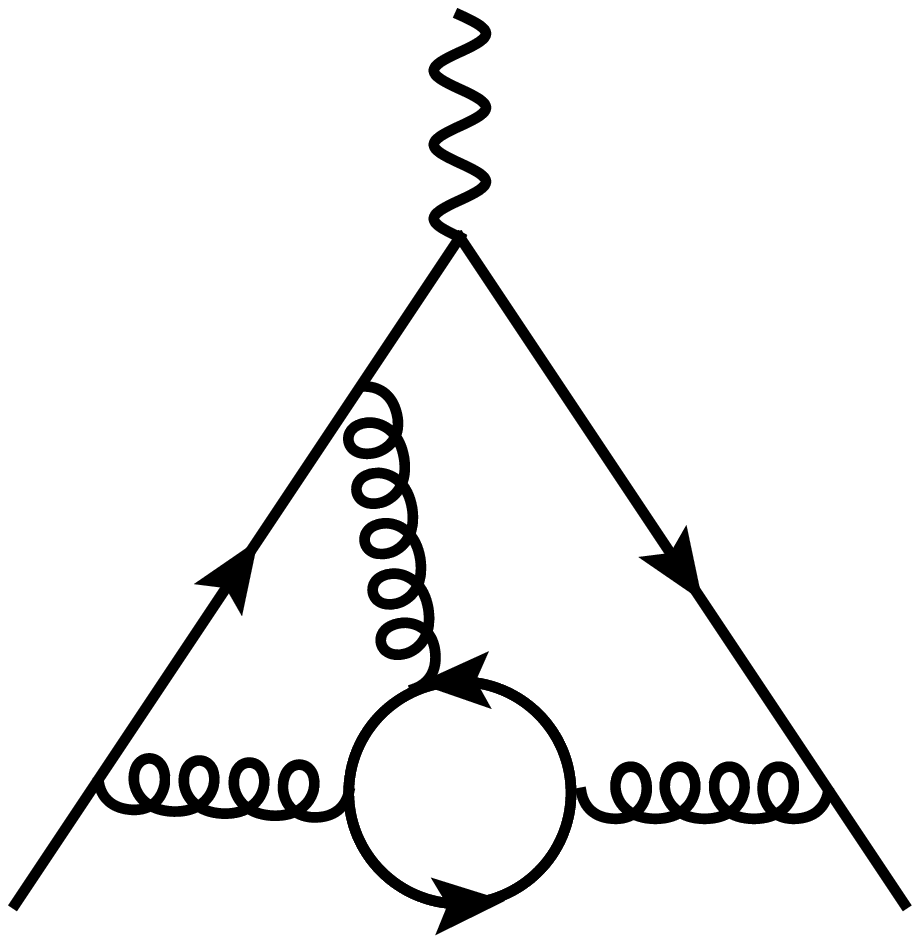}
      \\
      (a) & (b) & (c) & (d) & (e)
    \end{tabular}
    \caption{\label{fig::diags3}Sample diagrams contributing to $F_1$ and $F_2$
      at three-loop order. Solid, curly and wavy lines represent
      quarks, gluons and photons, respectively. In our calculation we
        only consider contributions with at least one closed massless quark loop.}
  \end{center}
\end{figure}

In the following we concentrate on the contributions to $F_1$ and $F_2$ which
contain at least one closed massless quark loop. Altogether there are 42 such
vertex diagrams, 41 of them contain exactly one closed massless fermion loop
and there is one diagram with two such closed loops.  Sample Feynman diagrams
contribution at three-loop order to the photon quark vertex are shown in
Fig.~\ref{fig::diags3}.

Note that some of the contributing planar diagrams are already present in the
large-$N_c$ limit~\cite{Henn:2016tyf} (see, e.g.,
Fig.~\ref{fig::diags3}(a)). However, other planar diagrams do not contribute
to the leading $N_c$ term and thus the corresponding integral families have
not been studied in Ref.~\cite{Henn:2016kjz}.  For example, the amplitude of
Fig.~\ref{fig::diags3}(b) is proportional to $C_F-C_A/2 =
1/(2N_c)$. Furthermore, there are non-planar contributions
(cf. Fig.~\ref{fig::diags3}(d)); all of them are sub-leading in the colour
factor and are treated for the first time in this paper.

\begin{figure}[t] 
  \begin{center}
    \begin{tabular}{cccc}
    \includegraphics[width=0.2\textwidth]{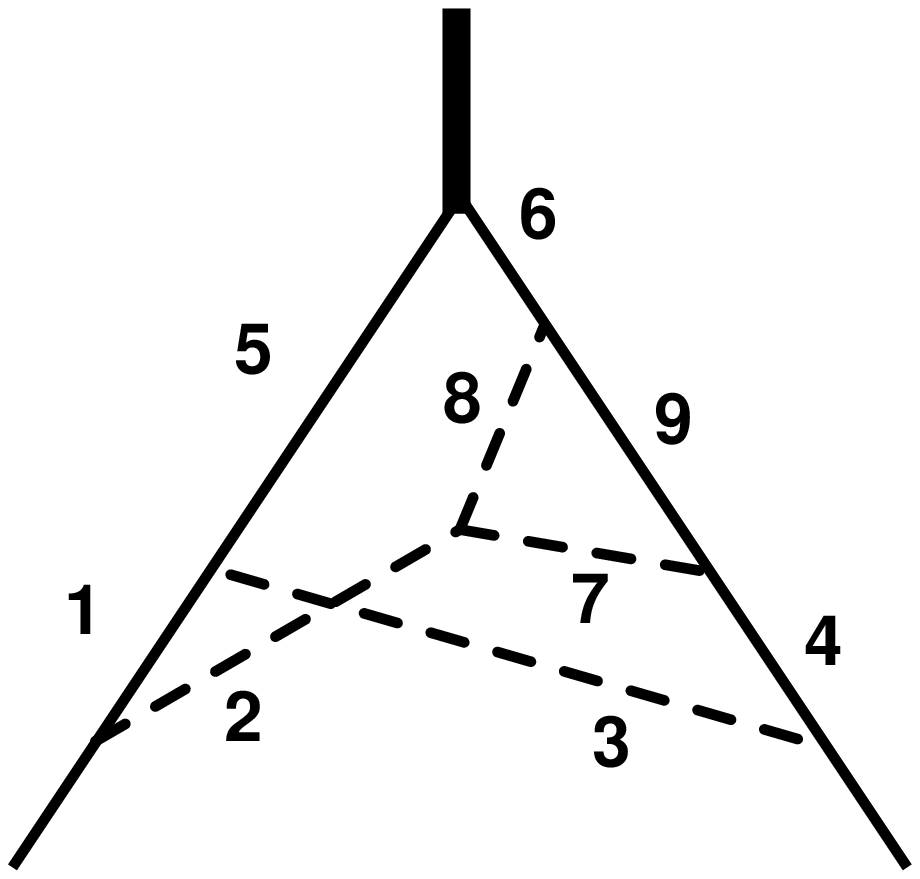}
    &
    \includegraphics[width=0.2\textwidth]{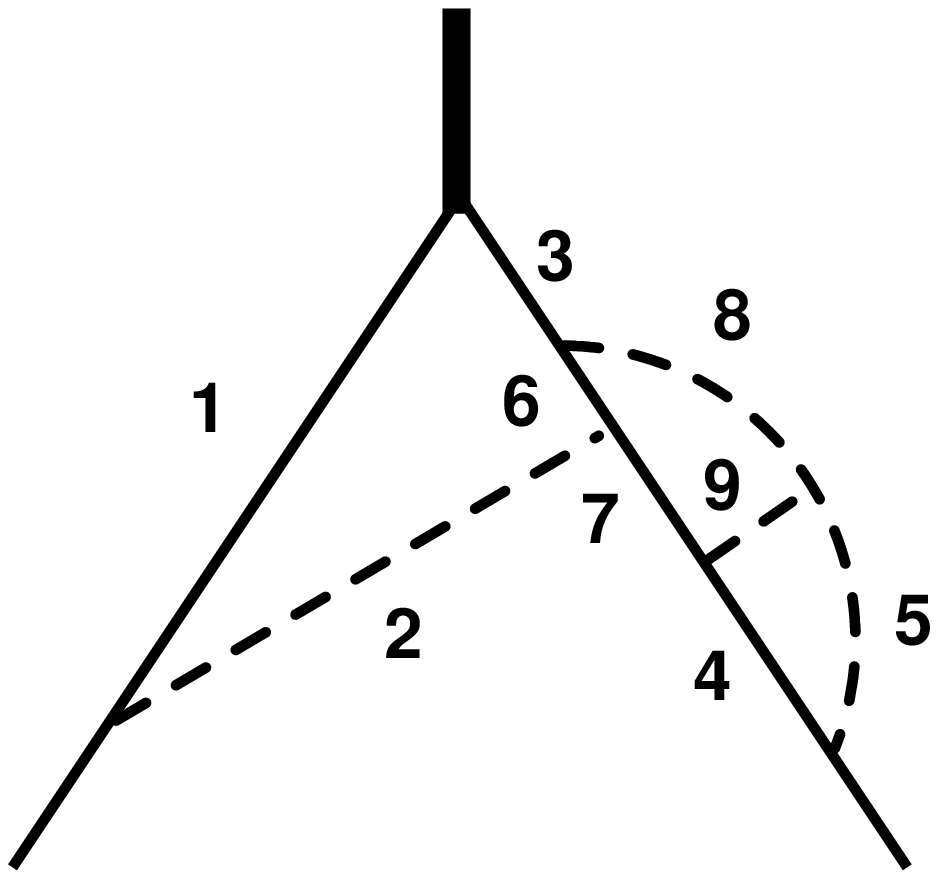}
    &
    \includegraphics[width=0.2\textwidth]{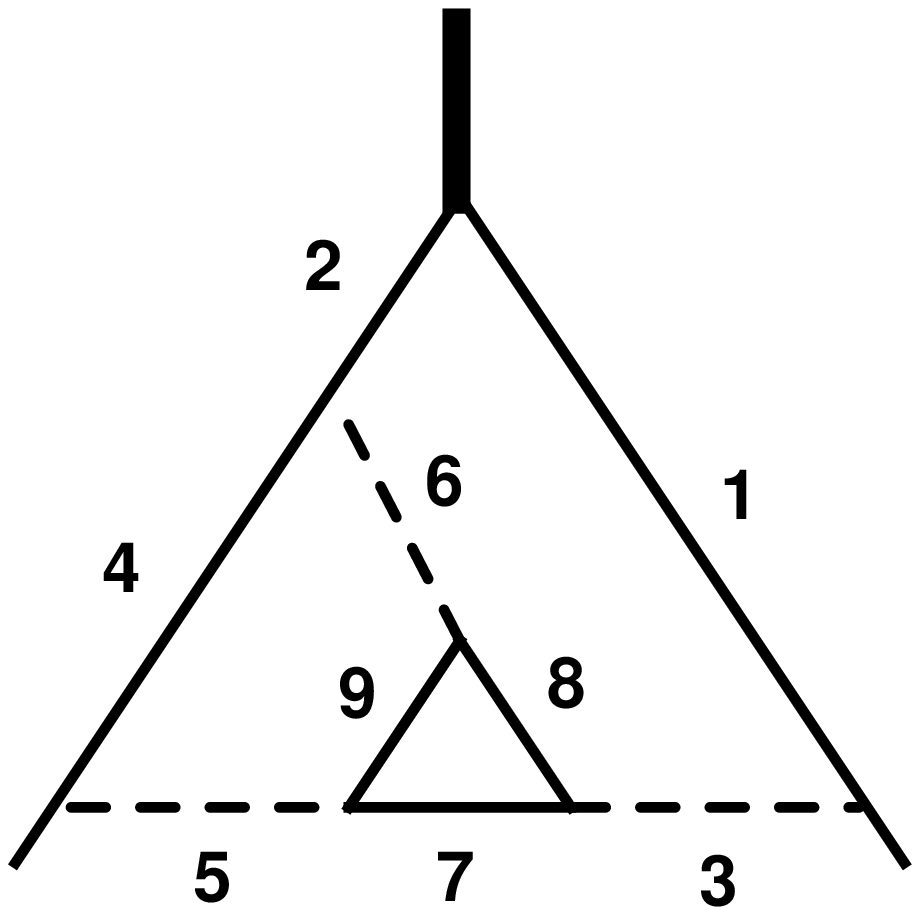}
    &
    \includegraphics[width=0.2\textwidth]{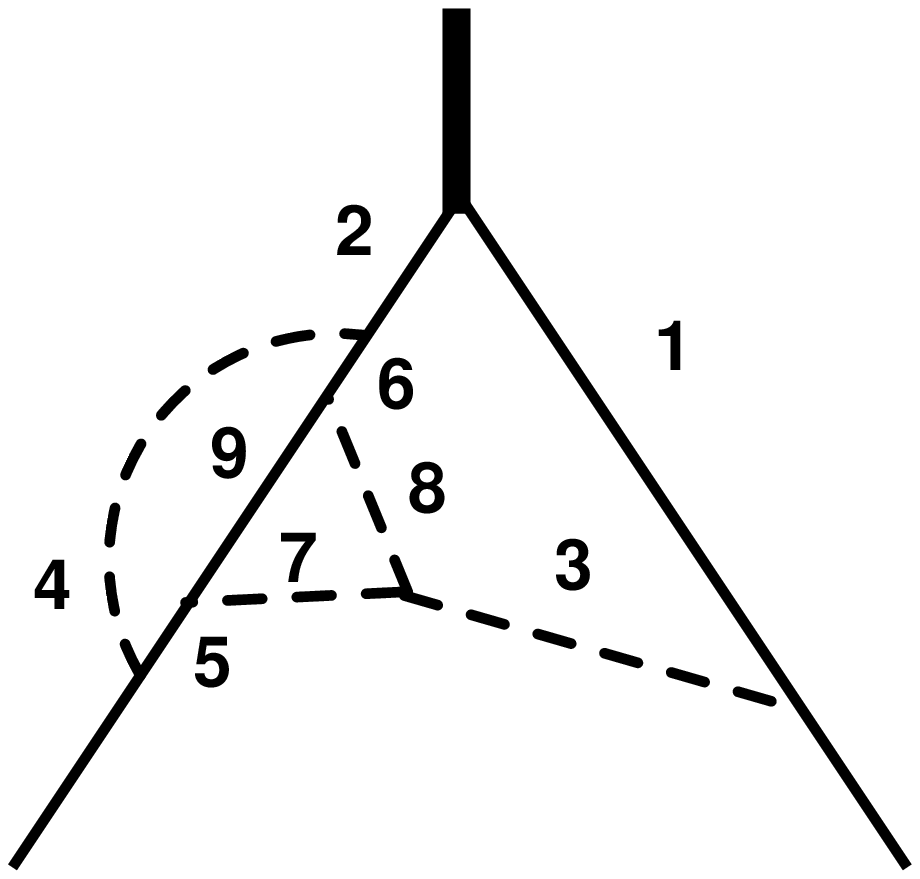}
    \\
    1051 & 1104 & 1136 & 1147
    \end{tabular}
    \caption{\label{fig::new_fams}New three-loop integral families
      needed for the fermionic contributions to the three-loop vertex
      corrections. Solid and dashed lines represent massive and
      massless lines, respectively.  Thin external lines are on the
      mass shell and thick external lines carry the off-shell momentum
      $q$.  For convenience we keep our internal numeration of
      the integral families, which is shown below the Feynman diagrams.}
  \end{center}
\end{figure}

For the three-loop calculation we define ten integral
families which are implemented in {\tt FIRE} and {\tt LiteRed}.
Six of them can be taken over from the large-$N_c$
calculation~\cite{Henn:2016kjz,Henn:2016tyf} and four are new.
Three of the new families are planar and one is non-planar,
see Fig.~\ref{fig::new_fams}.

To obtain results for the form factors we proceed as follows:
\begin{itemize}
\item We generate the amplitude for each diagram using {\tt
    qgraf}~\cite{Nogueira:1991ex} and transform the output to {\tt
    FORM}~\cite{Kuipers:2012rf} notation using {\tt q2e} and {\tt
    exp}~\cite{Harlander:1997zb,Seidensticker:1999bb}. The latter is also used
  to identify for each diagram the corresponding family and to perform the
  mapping of the integration momenta.
\item
  In a next step {\tt FORM} is used to evaluate the Dirac
  algebra. We apply the projectors to $F_1$ and $F_2$, perform the
  traces and decompose the scalar products, which appear in the numerator, to
  factors, which are present in the definition of the corresponding
  integral family. At this point each integral
  can be represented as a function which has the powers of the
  individual propagator factors as arguments.
  The list of integrals serves as input for {\tt
    FIRE}~\cite{Smirnov:2014hma}.  Note that we perform the calculation for
  general QCD gauge parameter $\xi$. $F_1$ and $F_2$ have to be
  independent of $\xi$ which serves as a welcome check for our
  calculation.
\item
  We use {\tt FIRE}~\cite{Smirnov:2014hma} in combination with {\tt
    LiteRed}~\cite{Lee:2012cn,Lee:2013mka} to generate integral tables for the ten
  families.  For the non-planar family, which is among the most complicated
  ones, this takes of the order of a week CPU time
  on a computer with about 100 GB RAM.
\item 
  Afterwards we minimize the set of the master integrals with the help of {\tt
    tsort}, which is part of the latest {\tt FIRE}
  version~\cite{Smirnov:2014hma} (implemented in the command {\tt
    FindRules}). It is based on ideas presented in
  Ref.~\cite{Smirnov:2013dia}, to obtain relations between primary master
  integrals, and to arrive at a minimal set.  Next we derive a system of
  differential equations for the master integrals using {\tt LiteRed}. We use
  {\tt FIRE} to reduce integrals appearing on the right-hand side of the
  equations.
\item
  In a next step we transform the system to $\epsilon$-form following
  the algorithm described in Ref.~\cite{Lee:2014ioa}.
\item 
  Our final result can be expressed in terms of GPLs with letters $-1,0$
  and $+1$, which is equivalent to Harmonic Polylogarithms (HPLs)~\cite{Remiddi:1999ew}.  Still we prefer
  to work with results in terms of GPLs, in particular, when taking various
  limits, because we use the same setup as in
  Refs.~\cite{Henn:2016kjz,Henn:2016tyf}.  Furthermore, in the calculation of
  the non-fermionic contributions to the massive form factor it will not be
  possible to express the result in terms of HPLs (see also
  Refs.~\cite{Henn:2016kjz,Henn:2016tyf}).
  
\item
  We consider the limit $q^2\to0$ to fix the boundary conditions. In
  this limit the vertex integrals become two-point on-shell integrals
  which are well-studied at three-loop order. We take the results
  from Ref.~\cite{Lee:2010ik}.
\end{itemize}

Results for all 89 planar master integrals entering the large-$N_c$
expressions for the form factors have been discussed in
Ref.~\cite{Henn:2016kjz} and explicit results have been presented.  In an
ancillary file to this paper~\cite{progdata} we present results for all master
integrals entering our results, which are not considered in
Ref.~\cite{Henn:2016kjz}. 
After minimizing the master integrals
of the four new families we observe that all integrals from family 1136 can be
mapped either to 1104 or 1147 or to the planar families studied in
Ref.~\cite{Henn:2016kjz} and we have to compute 15 new three-loop master
integrals from three families to obtain the results presented in this
paper.\footnote{Note that not all master integrals which are present in a
given family enter our result.}
 They are given by
\begin{eqnarray}
  &&G_{1051}(0, 0, 1, 0, 1, 1, 1, 2, 1, 0, 0, 0),\,\,
  G_{1051}(1, 0, 0, 1, 2, 0, 1, 1, 0, 0, 0, 0),\,\,
  \nonumber\\
  &&G_{1051}(1, 0, 1, 1, 1, 1, 1, 1, 0, 0, 0, 0),\,\,
  G_{1051}(1, 0, 1, 1, 1, 1, 1, 2, 0, 0, 0, 0),\,\,
  \nonumber\\
  &&G_{1104}(1, 0, 0, 0, 1, 1, 1, 0, 1, 0, 0, 0),\,\,
  G_{1104}(1, 0, 0, 0, 1, 1, 1, 0, 2, 0, 0, 0),\,\,
  \nonumber\\
  &&G_{1104}(1, 0, 0, 0, 1, 1, 2, 0, 1, 0, 0, 0),\,\,
  G_{1104}(1, 0, 1, 0, 1, 1, 1, 0, 1, 0, 0, 0),\,\,
  \nonumber\\
  &&G_{1104}(1, 0, 1, 0, 1, 1, 1, 0, 2, 0, 0, 0),\,\,
  G_{1147}(0, 1, 0, 0, 1, 1, 1, 1, 0, 0, 0, 0),\,\,
  \nonumber\\
  &&G_{1147}(1, 0, 0, 1, 1, 1, 1, 1, 0, 0, 0, 0),\,\,
  G_{1147}(1, 0, 0, 1, 1, 1, 1, 2, 0, 0, 0, 0),\,\,
  \nonumber\\
  &&G_{1147}(1, 0, 0, 1, 1, 2, 1, 1, 0, 0, 0, 0),\,\,
  G_{1147}(1, 1, 0, 0, 1, 1, 1, 1, 0, 0, 0, 0),\,\,
  \nonumber\\
  &&G_{1147}(1, 1, 0, 0, 1, 1, 1, 2, 0, 0, 0, 0)
  \,,
  \label{eq::3lMIs}
\end{eqnarray}
where the order of the indices corresponds to the line numbers introduced in
Fig.~\ref{fig::new_fams} and thus it is straightforward to construct the
integrands. Note that the last three indices represent irreducible numerators.
Since they are zero for all our integrals their precise definition is
irrelevant and we refrain from specifying them.  For all integrals in
Eq.~(\ref{eq::3lMIs}) we provide explicit results in~\cite{progdata}.  We
assume an integration measure $e^{\epsilon\gamma_E}{\rm d}^D k/(i\pi)^{D/2}$
with $D=4-2\epsilon$ and scalar propagators of the form $1/(m^2-k^2)$ or
$1/(-k^2)$.
Note that the above list only contains two non-planar integrals, $G_{1051}(1,
0, 1, 1, 1, 1, 1, 1, 0, 0, 0, 0)$ and $G_{1051}(1, 0, 1, 1, 1, 1, 1, 2, 0, 0,
0, 0)$.


\section{\label{sec::num}Analytical and numerical results}

In this section we discuss the results for the form factors $F_1$ and
$F_2$. The analytic results expressed in terms of GPLs are quite long and
we only present them in electronic form~\cite{progdata}. As already
mentioned above, they do not constitute physical results and in general still
contain poles in $\epsilon$.  Thus, we exemplify the numerical
results by considering the $\epsilon$-independent Taylor
coefficient.

We start with discussing analytic results in the low- and high-energy
and the threshold limit. 
The corresponding analytic expressions are also contained in an ancillary
file to this paper~\cite{progdata}.
They are obtained by expanding the Goncharov
polylogarithms of the exact result in the relevant limits.
Afterwards we demonstrate in Subsection~\ref{sub::num}
that a simple numerical evaluation of  $F_1$ and $F_2$
is possible.


\subsection{Form factors in the static limit}

In the static limit the form factors are infrared finite and thus
$F_1$ and $F_2$ do not contain poles in $\epsilon$. In the on-shell
scheme $F_1(q^2=0)=1$ and $F_2(q^2=0)$ is related to the quark
anomalous magnetic moment which we use as a cross check.  Note
that we use the limit $q^2\to0$ to fix the boundary conditions for the
master integrals (see discussion in Section~\ref{sec::3lres}).
However, this only requires as input scalar three-loop two-point on-shell
integrals (see Ref.~\cite{Lee:2010ik}) and thus the limit of the
final analytic expression for the form factor can still be
used as cross check. In fact, our explicit calculation shows that
$F_1(q^2=0)=1$ and $F_2(q^2=0)$ agrees with the dedicated three-loop calculation from
Ref.~\cite{Grozin:2007fh}.

We computed $F_1$ and $F_2$ up to order $(1-x)^6$ and refer to the ancillary
file for the complete expressions.  In the following we present results
for $F_1$ and $F_2$ up to ${\cal O}(\phi^2)$ including the constant
term in $\epsilon$. To obtain a manifest expansion $q^2\to0$ for
$q^2>0$ we use the variable $x=e^{i\phi}$ and display terms up to order
$\phi^2$. For $\mu^2=m^2$ we obtain the following results for $F_1$
\begin{eqnarray}
  F_1^{(1)} &=&
\phi^2\cR
\Bigg[
-\frac{2}{3 \ep}-\frac{1}{2}
\Bigg]
+{\cal O}(\epsilon)
  \,,
  \nonumber\\
  F_1^{(2)} &=&
\phi^2\Bigg\{\cR^2
\Bigg[
-12 \zeta (3)-\frac{47}{36}-\frac{175 \pi ^2}{54}+8 \pi ^2 \logtwo
\Bigg]
\nonumber\\&&\mbox{}
+\cA \cR
\Bigg[
\frac{11}{9 \ep^2}+\frac{\frac{2 \pi ^2}{9}-\frac{94}{27}}{\ep}+\frac{26 \zeta (3)}{3}+\frac{155 \pi ^2}{108}-\frac{2579}{324}-4 \pi ^2 \logtwo
\Bigg]
\nonumber\\&&\mbox{}
+\cR \tf \nl
\Bigg[
-\frac{4}{9 \ep^2}+\frac{20}{27 \ep}+\frac{8 \pi ^2}{27}+\frac{283}{81}
\Bigg]
+\cR \tf \nh
\Bigg[
\frac{3 \pi ^2}{2}-\frac{1099}{81}
\Bigg]
\Bigg\}
\nonumber\\&&\mbox{} 
+{\cal O}(\epsilon)
  \,,
  \nonumber\\
  F_1^{(3)}\Big|_{n_l} &=&
\phi^2\Bigg\{\cR^2 \tf \nl
\Bigg[
-\frac{8}{9 \ep^2}
+\frac{\frac{110}{27}-\frac{32 \zthree}{9}}{\ep}
+\frac{512 \afour}{3}
+\frac{64 \logtwo^4}{9}+\frac{128 \pi ^2 \logtwo^2}{9}
\nonumber\\&&\mbox{}
-\frac{1768 \pi ^2
  \logtwo}{27}
+\frac{1100 \zthree}{9}-\frac{448 \pi ^4}{135}+\frac{9838 \pi ^2}{243}-\frac{3107}{162}
\Bigg]
\nonumber\\&&\mbox{}
+\cA \cR \tf \nl
\Bigg[
\frac{176}{81 \ep^3}
+\frac{\frac{16 \pi ^2}{81}-\frac{1552}{243}}{\ep^2}
+\frac{\frac{112 \zthree}{27}-\frac{160 \pi ^2}{243}+\frac{1556}{243}}{\ep}
-\frac{256 \afour}{3}
\nonumber\\&&\mbox{}
-\frac{32 \logtwo^4}{9}-\frac{64 \pi ^2 \logtwo^2}{9}+\frac{884 \pi ^2 \logtwo}{27}-\frac{1622 \zthree}{27}+\frac{352 \pi ^4}{405}-\frac{5237 \pi ^2}{729}+\frac{260644}{2187}
\Bigg]
\nonumber\\&&\mbox{}
+\cR \tf^2 \nl^2
\Bigg[
-\frac{32}{81 \ep^3}+\frac{160}{243 \ep^2}+\frac{32}{243 \ep}-\frac{448 \zthree}{81}-\frac{464 \pi ^2}{243}-\frac{29524}{2187}
\Bigg]
\nonumber\\&&\mbox{}
+\cR \tf^2 \nh \nl
\Bigg[
\frac{8 \pi ^2}{81 \ep}-\frac{8 \pi ^2 \logtwo}{3}+\frac{724 \zthree}{81}-\frac{892 \pi ^2}{243}+\frac{10088}{243}
\Bigg]
\Bigg\}
+{\cal O}(\epsilon)
  \,,
\end{eqnarray}
where $\logtwo=\log(2)$ and $\afour=\mbox{Li}_4(1/2)$.
For $F_2$ we have
\begin{eqnarray}
  F_2^{(1)} &=&
  2\cR+ \phi^2 \frac{\cR}{3}
+{\cal O}(\epsilon)  \,,
\nonumber\\
F_2^{(2)} &=& 
\cR^2
\Bigg[
-8 \pi ^2 \logtwo+12 \zthree+\frac{20 \pi ^2}{3}-31
\Bigg]
+
\cA \cR
\Bigg[
4 \pi ^2 \logtwo-6 \zthree-2 \pi ^2+\frac{317}{9}
\Bigg]
\nonumber\\&&\mbox{}
+\cR \tf \nl
\Bigg[
-\frac{100}{9}
\Bigg]
+ \cR \tf \nh
\Bigg[
\frac{476}{9}-\frac{16 \pi ^2}{3}
\Bigg]
+\phi^2\Bigg\{
\cR^2
\Bigg[
  -\frac{4}{3 \ep}-\frac{92 \pi ^2 \logtwo}{15}
\nonumber\\&&\mbox{}
  +\frac{46 \zthree}{5}+\frac{61 \pi ^2}{15}-\frac{77}{5}
\Bigg]
+ \cA \cR
\Bigg[
  \frac{46 \pi ^2 \logtwo}{15}-\frac{23 \zthree}{5}
  -\frac{137 \pi ^2}{90}+\frac{1699}{270}
\Bigg]
\nonumber\\&&\mbox{}
+ \cR \tf \nl
\Bigg[
-\frac{62}{27}
\Bigg]
+ \cR \tf \nh
\Bigg[
\frac{622}{27}-\frac{7 \pi ^2}{3}
\Bigg]
\Bigg\}
+{\cal O}(\epsilon)\,,
\nonumber\\
F_2^{(3)}\Big|_{n_l} &=& 
\cR^2 \tf \nl
\Bigg[
-\frac{512 \afour}{3}-\frac{64 \logtwo^4}{9}-\frac{128 \pi ^2
  \logtwo^2}{9}+\frac{320 \pi ^2 \logtwo}{3}-192 \zthree+\frac{88 \pi
  ^4}{27}-\frac{2528 \pi ^2}{27}
\nonumber\\&&\mbox{}
+250
\Bigg]
+
\cA \cR \tf \nl
\Bigg[
\frac{256 \afour}{3}+\frac{32 \logtwo^4}{9}+\frac{64 \pi ^2
  \logtwo^2}{9}-\frac{160 \pi ^2 \logtwo}{3}+\frac{304
  \zthree}{3}-\frac{44 \pi ^4}{27}
\nonumber\\&&\mbox{}
+\frac{616 \pi ^2}{27}-\frac{38576}{81}
\Bigg]
+\cR \tf^2 \nl^2
\Bigg[
\frac{5072}{81}+\frac{64 \pi ^2}{27}
\Bigg]
+\cR \tf^2 \nh \nl
\Bigg[
\frac{64 \pi ^2}{27}-\frac{1952}{81}
\Bigg]
\nonumber\\&&\mbox{}
+\phi^2\Bigg\{
\cR^2 \tf \nl
\Bigg[
-\frac{5888 \afour}{45}-\frac{8}{9 \ep^2}+\frac{16}{3 \ep}-\frac{736
  \logtwo^4}{135}-\frac{1472 \pi ^2 \logtwo^2}{135}+\frac{8048 \pi ^2
  \logtwo}{135}
\nonumber\\&&\mbox{}
-\frac{664 \zthree}{5}+\frac{1012 \pi ^4}{405}-\frac{6092 \pi ^2}{135}+\frac{12653}{90}
\Bigg]
+\cA \cR \tf \nl
\Bigg[
  \frac{2944 \afour}{45}+\frac{368 \logtwo^4}{135}
\nonumber\\&&\mbox{}
  +\frac{736 \pi ^2 \logtwo^2}{135}-\frac{4024 \pi ^2 \logtwo}{135}+\frac{348 \zthree}{5}-\frac{506 \pi ^4}{405}+\frac{458 \pi ^2}{27}-\frac{26626}{243}
\Bigg]
\nonumber\\&&\mbox{}
+\cR \tf^2 \nl^2
\Bigg[
\frac{3736}{243}+\frac{32 \pi ^2}{81}
\Bigg]
+\cR \tf^2 \nh \nl
\Bigg[
  \frac{16 \pi ^2 \logtwo}{9}-\frac{56 \zthree}{9}+\frac{40 \pi ^2}{9}
\nonumber\\&&\mbox{}
  -\frac{11824}{243}
\Bigg]
\Bigg\}
+{\cal O}(\epsilon)\,.
\end{eqnarray}
Note that starting from the next-to-leading expansion term of order
$\phi^2$ both $F_1$ and $F_2$ are infrared divergent and develops
poles in $\epsilon$.


\subsection{\label{sub::he}Form factors at high energies}

In the limit $x\to0$ we compute terms up to ${\cal O}(x^6)$ both for
$F_1$ and $F_2$. To illustrate the structure of the analytic
expressions we show the first two terms of order $x^0$ and $x^1$ for
$F_1$ at three loops.
After introducing the notation
\begin{eqnarray}
  F_i^{(n)} &=& \sum_{k\ge0} f_{i,\rm lar}^{(n,k)} x^k
  \,.
  \label{eq::F_i_lar}
\end{eqnarray}
we have
\begin{eqnarray}
  f_{1,\rm lar}^{(1,0)} &=&
\cR
\Bigg[
\left(-\frac{2}{\ep}-3\right)
 \logx-\frac{2}{\ep}-\logx^2+\frac{\pi ^2}{3}-4
\Bigg]
  \,,\nonumber\\
  f_{1,\rm lar}^{(1,1)} &=&
\cR
\Bigg[
2 \logx-4
\Bigg]
  \,,\nonumber\\
  f_{1,\rm lar}^{(2,0)} &=&
\cR^2
\Bigg[
\left(\frac{2}{\ep^2}+\frac{8}{\ep}-\frac{2 \pi ^2}{3}+\frac{55}{2}\right)
 \logx^2+\logx \left(\frac{4}{\ep^2}+\frac{14-\frac{2 \pi ^2}{3}}{\ep}-32 \zeta (3)+\frac{85}{2}\right)
+\frac{2}{\ep^2}
\nonumber\\&&\mbox{}
+\left(\frac{2}{\ep}+\frac{20}{3}\right)
 \logx^3+\frac{8-\frac{2 \pi ^2}{3}}{\ep}+\frac{7 \logx^4}{6}-44 \zeta (3)-\frac{59 \pi ^4}{90}+\frac{13 \pi ^2}{2}+46-8 \pi ^2 \logtwo
\Bigg]
\nonumber\\&&\mbox{}
+ \cA \cR
\Bigg[
\logx \left(\frac{11}{3 \ep^2}+\frac{\frac{\pi ^2}{3}-\frac{67}{9}}{\ep}+26 \zeta (3)-\frac{11 \pi ^2}{9}-\frac{2545}{54}\right)
\nonumber\\&&\mbox{}
+\frac{11}{3 \ep^2}+\frac{-2 \zeta (3)-\frac{49}{9}+\frac{\pi ^2}{3}}{\ep}-\frac{11 \logx^3}{9}+\left(\frac{\pi ^2}{3}-\frac{233}{18}\right)
 \logx^2
\nonumber\\&&\mbox{}
+\frac{134 \zeta (3)}{3}-\frac{\pi ^4}{60}-\frac{7 \pi ^2}{54}-\frac{1595}{27}+4 \pi ^2 \logtwo
\Bigg]
\nonumber\\&&\mbox{}
+ \cR \tf \nl
\Bigg[
\left(-\frac{4}{3 \ep^2}+\frac{20}{9 \ep}+\frac{4 \pi ^2}{9}+\frac{418}{27}\right)
 \logx
\nonumber\\&&\mbox{}
-\frac{4}{3 \ep^2}+\frac{20}{9 \ep}+\frac{4 \logx^3}{9}+\frac{38 \logx^2}{9}-\frac{16 \zeta (3)}{3}-\frac{14 \pi ^2}{27}+\frac{424}{27}
\Bigg]
\nonumber\\&&\mbox{}
+ \cR \tf \nh
\Bigg[
\frac{4 \logx^3}{9}+\frac{38 \logx^2}{9}+\left(\frac{530}{27}+\frac{2 \pi ^2}{3}\right)
 \logx-\frac{4 \pi ^2}{9}+\frac{1532}{27}
\Bigg]
  \,,\nonumber\\
  f_{1,\rm lar}^{(2,1)} &=&
\cR^2
\Bigg[
\left(-\frac{4}{\ep}+\frac{4 \pi ^2}{3}-37\right)
 \logx^2+\logx \left(\frac{4}{\ep}-48 \zeta (3)+6 \pi ^2+13\right)
\nonumber\\&&\mbox{}
+\frac{8}{\ep}-\frac{\logx^4}{3}-\frac{28 \logx^3}{3}-88 \zeta (3)+\frac{32 \pi ^4}{45}-5 \pi ^2-22+48 \pi ^2 \logtwo
\Bigg]
\nonumber\\&&\mbox{}
+ \cA \cR
\Bigg[
\frac{\logx^4}{6}+\frac{8 \logx^3}{3}+\left(\frac{4 \pi ^2}{3}-\frac{25}{3}\right)
 \logx^2+\logx \left(-72 \zeta (3)+\frac{341}{9}+\frac{22 \pi ^2}{3}\right)
\nonumber\\&&\mbox{}
-200 \zeta (3)+\frac{7 \pi ^4}{9}+\frac{247 \pi ^2}{9}-\frac{904}{9}-24 \pi ^2 \logtwo
\Bigg]
\nonumber\\&&\mbox{}
+\cR \tf \nl
\Bigg[
-\frac{4 \logx^2}{3}-\frac{148 \logx}{9}+\frac{4 \pi ^2}{9}+\frac{200}{9}
\Bigg]
\nonumber\\&&\mbox{}
+\cR \tf \nh
\Bigg[
-\frac{52 \logx^2}{3}-\frac{436 \logx}{9}-\frac{44 \pi ^2}{3}-\frac{784}{9}
\Bigg]
  \,,\nonumber\\
  f_{1,\rm lar}^{(3,0)}\Big|_{n_l} &=&
\cR^2 \tf \nl
\Bigg[
\frac{8}{3 \ep^3}+\left(\frac{4}{3 \ep^2}-\frac{82}{9 \ep}-\frac{29 \pi ^2}{27}-\frac{2032}{27}\right)
 \logx^3+\frac{-\frac{16}{9}-\frac{4 \pi ^2}{9}}{\ep^2}
\nonumber\\&&\mbox{}
+\logx^2 \left(\frac{8}{3 \ep^3}+\frac{8}{9 \ep^2}+\frac{-\frac{962}{27}-\frac{10 \pi ^2}{9}}{\ep}+\frac{232 \zeta (3)}{9}-\frac{50 \pi ^2}{9}-\frac{18817}{81}\right)
\nonumber\\&&\mbox{}
+\logx \left(\frac{16}{3 \ep^3}+\frac{-\frac{20}{9}-\frac{4 \pi ^2}{9}}{\ep^2}+\frac{-\frac{16 \zeta (3)}{3}-\frac{1198}{27}+\frac{2 \pi ^2}{9}}{\ep}+\frac{1976 \zeta (3)}{9}+\frac{98 \pi ^4}{135}-\frac{341 \pi ^2}{27}
\right.\nonumber\\&&\left.\mbox{}
-\frac{18812}{81}\right)
+\left(-\frac{4}{9 \ep}-\frac{355}{27}\right)
 \logx^4+\frac{-\frac{16 \zeta (3)}{3}-\frac{470}{27}+\frac{4 \pi ^2}{3}}{\ep}-\logx^5-\frac{512}{3} \afour
\nonumber\\&&\mbox{}
+40 \zeta (5)-8 \pi ^2 \zeta (3)+\frac{2752 \zeta (3)}{9}+\frac{3058 \pi
  ^4}{405}-\frac{481 \pi ^2}{9}-\frac{2011}{81}-\frac{64 \logtwo^4}{9}
\nonumber\\&&\mbox{}
-\frac{128}{9} \pi ^2 \logtwo^2+\frac{224}{9} \pi ^2 \logtwo
\Bigg]
\nonumber\\&&\mbox{}
+ \cA \cR \tf \nl
\Bigg[
\frac{176}{27 \ep^3}+\frac{-\frac{16 \zeta
    (3)}{9}-\frac{1192}{81}+\frac{8 \pi ^2}{27}}{\ep^2}
+\logx \left(\frac{176}{27 \ep^3}+\frac{\frac{8 \pi
      ^2}{27}-\frac{1336}{81}}{\ep^2}
\right.\nonumber\\&&\left.\mbox{}
+\frac{\frac{112 \zeta (3)}{9}+\frac{836}{81}-\frac{80 \pi ^2}{81}}{\ep}-\frac{1448 \zeta (3)}{9}-\frac{22 \pi ^4}{135}+\frac{5864 \pi ^2}{243}+\frac{309838}{729}\right)
\nonumber\\&&\mbox{}
+\frac{\frac{496 \zeta (3)}{27}+\frac{356}{81}-\frac{80 \pi ^2}{81}}{\ep}+\frac{44 \logx^4}{27}+\left(\frac{1948}{81}-\frac{8 \pi ^2}{27}\right)
 \logx^3+\logx^2 \left(-16 \zeta (3)+\frac{11752}{81}
\right.\nonumber\\&&\left.\mbox{}
+\frac{16 \pi ^2}{9}\right)
+\frac{256}{3} \afour
+\frac{596 \zeta (5)}{3}+\frac{4 \pi ^2 \zeta (3)}{9}-\frac{31120 \zeta
  (3)}{81}-\frac{1822 \pi ^4}{405}
\nonumber\\&&\mbox{}
+\frac{1504 \pi ^2}{243}+\frac{259150}{729}
+\frac{32 \logtwo^4}{9}+\frac{64}{9} \pi ^2 \logtwo^2-\frac{112}{9} \pi ^2 \logtwo
\Bigg]
\nonumber\\&&\mbox{}
+\cR \tf^2 \nl^2
\Bigg[
-\frac{32}{27 \ep^3}+\frac{160}{81 \ep^2}+\logx \left(-\frac{32}{27
    \ep^3}+\frac{160}{81 \ep^2}+\frac{32}{81 \ep}-\frac{64 \zeta
    (3)}{27}-\frac{304 \pi ^2}{81}
\right.\nonumber\\&&\left.\mbox{}
-\frac{39352}{729}\right)
+\frac{32}{81 \ep}-\frac{8 \logx^4}{27}-\frac{304 \logx^3}{81}+\left(-\frac{1624}{81}-\frac{16 \pi ^2}{27}\right)
 \logx^2+\frac{256 \zeta (3)}{9}+\frac{232 \pi ^4}{405}
\nonumber\\&&\mbox{}
-\frac{488 \pi ^2}{243}-\frac{29344}{729}
\Bigg]
\nonumber\\&&\mbox{}
+\cR \tf^2 \nh \nl
\Bigg[
\logx \left(\frac{8 \pi ^2}{27 \ep}-\frac{416 \zeta (3)}{27}-8 \pi ^2-\frac{7408}{81}\right)
+\frac{8 \pi ^2}{27 \ep}-\frac{16 \logx^4}{27}-\frac{608 \logx^3}{81}
\nonumber\\&&\mbox{}
+\left(-\frac{3248}{81}-\frac{32 \pi ^2}{27}\right)
 \logx^2
-\frac{416 \zeta (3)}{9}-\frac{16 \pi ^4}{27}+\frac{776 \pi ^2}{243}-\frac{5072}{27}
\Bigg]
  \,,\nonumber\\
  { f_{1,\rm lar}^{(3,1)}\Big|_{n_l} } &=&
\cR^2 \tf \nl
\Bigg[
\logx^2 \left(-\frac{8}{3 \ep^2}+\frac{92}{3 \ep}+\frac{16 \zeta (3)}{3}-\frac{178 \pi ^2}{27}+\frac{3416}{9}\right)
\nonumber\\&&\mbox{}
+\logx \left(\frac{8}{3 \ep^2}+\frac{-\frac{32}{3}-\frac{4 \pi ^2}{9}}{\ep}+\frac{224 \zeta (3)}{3}+\frac{344 \pi ^4}{135}-\frac{2918 \pi ^2}{27}+\frac{790}{9}\right)
\nonumber\\&&\mbox{}
+\frac{16}{3 \ep^2}+\left(\frac{4}{3 \ep}-\frac{8 \pi ^2}{3}+\frac{3436}{27}\right)
 \logx^3+\frac{-40-\frac{4 \pi ^2}{9}}{\ep}+\frac{4 \logx^5}{9}+\frac{352 \logx^4}{27}+1024 \afour
\nonumber\\&&\mbox{}
-\frac{1312 \zeta (5)}{3}+\frac{64 \pi ^2 \zeta (3)}{3}
+\frac{4528 \zeta (3)}{9}-\frac{524 \pi ^4}{405}+\frac{3400 \pi
  ^2}{27}-\frac{1676}{9}+\frac{128 \logtwo^4}{3}
\nonumber\\&&\mbox{}
+\frac{256}{3} \pi ^2 \logtwo^2-\frac{1664}{3} \pi ^2 \logtwo
\Bigg]
\nonumber\\&&\mbox{}
+\cA \cR \tf \nl
\Bigg[
-\frac{2 \logx^5}{9}-\frac{116 \logx^4}{27}+\left(-\frac{160}{27}-\frac{4 \pi ^2}{3}\right)
 \logx^3+\logx^2 \left(\frac{232 \zeta (3)}{3}+\frac{104}{27}
\right.\nonumber\\&&\left.\mbox{}
-\frac{520 \pi ^2}{27}\right)
+\logx \left(560 \zeta (3)-\frac{34664}{81}-\frac{788 \pi ^2}{9}+\frac{236 \pi ^4}{135}\right)
-512 \afour
\nonumber\\&&\mbox{}
-\frac{1840 \zeta (5)}{3}-\frac{32 \pi ^2 \zeta (3)}{3}
+\frac{14440 \zeta (3)}{9}+\frac{868 \pi ^4}{405}-\frac{16988 \pi
  ^2}{81}+\frac{97384}{81}
\nonumber\\&&\mbox{}
-\frac{64 \logtwo^4}{3}-\frac{128}{3} \pi ^2 \logtwo^2
+\frac{832}{3} \pi ^2 \logtwo
\Bigg]
\nonumber\\&&\mbox{}
+\cR \tf^2 \nl^2
\Bigg[
\frac{32 \logx^3}{27}+\frac{592 \logx^2}{27}+\left(\frac{8720}{81}+\frac{32 \pi ^2}{27}\right)
 \logx-\frac{128 \zeta (3)}{9}-\frac{976 \pi ^2}{81}-\frac{11296}{81}
\Bigg]
\nonumber\\&&\mbox{}
+\cR \tf^2 \nh \nl
\Bigg[
\frac{448 \logx^3}{27}+\frac{3104 \logx^2}{27}+\left(\frac{15136}{81}+\frac{448 \pi ^2}{27}\right)
 \logx+\frac{1664 \zeta (3)}{9}
\nonumber\\&&\mbox{}
+\frac{5696 \pi ^2}{81}-\frac{5408}{81}
\Bigg]
  \,,
\label{eq::fflar}
\end{eqnarray}
where $l_x = \log(x)$. 
It is interesting to note that at three-loop order $l_x$ may in principle
appear up to sixth power.
However, for the $n_l$ terms at most $l_x^5$ terms
are present. In the case of $f_{1,\rm lar}^{(3,0)}$
the $l_x^6$ term comes with the colour factor $C_F^3$ which is known since long
\cite{Sudakov:1954sw,Frenkel:1976bj}.
In Ref.~\cite{Penin:2014msa,Liu:2017axv,Liu:2017vkm} it has been shown 
that the
$l_x^6$ term in the power-suppressed contribution $f_{1,\rm lar}^{(3,1)}$
comes together with colour structures $C_F-C_A/2$ in the
$n_l$-independent term.  Explicit results for power-suppressed terms are given
in Ref.~\cite{Liu:2017axv}.

For $F_2$ we have $f_{2,\rm lar}^{(n,0)} =0$ (for
$n=1,2$ and $3$) and
\begin{eqnarray}
  f_{2,\rm lar}^{(1,1)} &=& -4\cR \logx
  \,,\nonumber\\
  f_{2,\rm lar}^{(2,1)} &=&
\cR^2
\Bigg[
\left(\frac{8}{\ep}+34\right)
 \logx^2+\left(\frac{8}{\ep}-8 \pi ^2+62\right)
 \logx-32 \pi ^2 \logtwo+8 \logx^3+16 \zthree+10 \pi ^2
\Bigg]
\nonumber\\&&\mbox{}
+\cA \cR
\Bigg[
16 \pi ^2 \logtwo+\frac{2 \logx^2}{3}-\frac{346 \logx}{9}+80 \zthree-\frac{122 \pi ^2}{9}+12
\Bigg]
\nonumber\\&&\mbox{}
+\cR \tf \nl
\Bigg[
\frac{8 \logx^2}{3}+\frac{200 \logx}{9}-\frac{8 \pi ^2}{9}
\Bigg]
+\cR \tf \nh
\Bigg[
\frac{8 \logx^2}{3}+\frac{200 \logx}{9}-\frac{8 \pi ^2}{3}+\frac{272}{3}
\Bigg]
  \,,\nonumber\\
  f_{2,\rm lar}^{(3,1)} &=&
\cR^2 \tf \nl
\Bigg[
-\frac{2048 \afour}{3}+\left(\frac{16}{3 \ep^2}-\frac{104}{3 \ep}-\frac{28 \pi ^2}{9}-\frac{2936}{9}\right)
 \logx^2
+\logx \left(\frac{16}{3 \ep^2}
\right.\nonumber\\&&\left.\mbox{}
+\frac{\frac{8 \pi ^2}{9}-32}{\ep}+\frac{416 \zthree}{3}+\frac{1588 \pi ^2}{27}-\frac{4172}{9}\right)
+\left(-\frac{8}{3 \ep}-\frac{832}{9}\right)
 \logx^3
+\frac{8 \pi ^2}{9 \ep}
\nonumber\\&&\mbox{}
-\frac{256 \logtwo^4}{9}-\frac{512 \pi ^2 \logtwo^2}{9}+\frac{2816 \pi ^2 \logtwo}{9}-\frac{80 \logx^4}{9}-\frac{3328 \zthree}{9}+\frac{656 \pi ^4}{135}-\frac{2720 \pi ^2}{27}+\frac{16}{3}
\Bigg]
\nonumber\\&&\mbox{}
+\cA \cR \tf \nl
\Bigg[
\frac{1024 \afour}{3}+\frac{128 \logtwo^4}{9}+\frac{256 \pi ^2
  \logtwo^2}{9}-\frac{1408 \pi ^2 \logtwo}{9}+\frac{64 \logx^3}{27}
+\left(\frac{1256}{27}
\right.\nonumber\\&&\left.\mbox{}
+\frac{40 \pi ^2}{9}\right)
 \logx^2+\logx \left(-\frac{512 \zthree}{3}+\frac{1168 \pi ^2}{27}+\frac{44320}{81}\right)
-\frac{2192 \zthree}{3}-\frac{8 \pi ^4}{3}
\nonumber\\&&\mbox{}
+\frac{8080 \pi ^2}{81}-\frac{496}{3}
\Bigg]
+ \cR \tf^2 \nl^2
\Bigg[
-\frac{64 \logx^3}{27}-\frac{800 \logx^2}{27}+\left(-\frac{10144}{81}-\frac{64
    \pi ^2}{27}\right) \logx
\nonumber\\&&\mbox{}
 +\frac{256 \zthree}{9}+\frac{800 \pi ^2}{81}
\Bigg]
+\cR \tf^2 \nh \nl
\Bigg[
-\frac{128 \logx^3}{27}-\frac{1600 \logx^2}{27}
\nonumber\\&&\mbox{}
+\left(-\frac{20288}{81}-\frac{128 \pi ^2}{27}\right)
 \logx-\frac{256 \zthree}{9}-\frac{3712 \pi ^2}{81}-\frac{2240}{27}
\Bigg]
  \,.
\label{eq::fflar_2}
\end{eqnarray}


\subsection{Form factors and threshold cross section}

In the threshold limit ($q^2\to 4 m^2$ or $x\to-1$) the form factors $F_1$ and
$F_2$ can be used to obtain the physical cross section for $e^+e^- \to
\mbox{hadrons}$ since they constitute the virtual corrections and the real
corrections are suppressed by a relative factor $\beta^3$.  This means
that we can predict $\sigma(e^+e^-\to Q \bar{Q})$ including terms of
order $\beta^{2-n}$ at $n$-loop order. Since for the
three-loop $n_l$ contribution the $\beta^0$ terms computed
from $F_1$ and $F_2$ are finite, we also show them below.

For convenience we
repeat the formula which can be used to compute the cross section from
the form factors (see also Ref.~\cite{Henn:2016tyf}) which reads
\begin{eqnarray}
  \sigma(e^+e^-\to Q \bar{Q}) &=& \sigma_0 \beta
  \left[ |F_1 + F_2|^2 
    + \frac{ |(1-\beta^2) F_1 + F_2 |^2}{2(1-\beta^2)}
  \right] 
  \nonumber\\
  &=& \sigma_0 \frac{3\beta}{2}\left[
    1-\frac{\beta^2}{3}
    + \frac{\alpha_s}{4\pi} \Delta^{(1)} 
    + \left(\frac{\alpha_s}{4\pi}\right)^2 \Delta^{(2)}
    + \left(\frac{\alpha_s}{4\pi}\right)^3 \Delta^{(3)}
    + \ldots
  \right]
  \,,
  \label{eq::sigee}
\end{eqnarray}
where $\sigma_0 = 4\pi \alpha^2 Q_Q^2 /(3s)$.  Using the results from this
paper we obtain complete expressions for $\Delta^{(1)}$ and $\Delta^{(2)}$ and all
$n_l$ terms for $\Delta^{(3)}$ which are given by
\begin{eqnarray}
  \Delta^{(1)} &=&
  \cR
  \Bigg[
     \frac{1}{\beta} 2 \pi ^2 -16+ \beta
    \Bigg(\frac{4 \pi ^2}{3}\Bigg)
    \Bigg]
    + \ldots
  \,, \nonumber\\
  \Delta^{(2)} &=&
\cR^2
\Bigg[
 \frac{4 \pi ^4}{3\beta^2} - \frac{1}{\beta} 32
\pi ^2 -\frac{32}{3} \pi ^2 \log (2 \beta )-16 \zeta (3)+\frac{20
  \pi ^4}{9}
\nonumber\\&&\mbox{}
-\frac{280 \pi ^2}{9}+156+32 \pi ^2 \logtwo
\Bigg]
+\cA \cR
\Bigg[
\frac{1}{\beta} \Bigg(\frac{62 \pi ^2}{9}-\frac{44}{3} \pi ^2 \log (2 \beta
)\Bigg)
\nonumber\\&&\mbox{}
-16 \pi ^2 \log (2 \beta )-104 \zeta (3)+\frac{358 \pi ^2}{9}
-\frac{604}{9}-\frac{80}{3} \pi ^2 \logtwo
\Bigg]
\nonumber\\&&\mbox{}
+\cR \tf \nl
\Bigg[
\frac{1}{\beta} \Bigg(\frac{16}{3} \pi ^2 \log (2 \beta )-\frac{40 \pi ^2}{9}\Bigg)+ \frac{176}{9}
\Bigg]
\nonumber\\&&\mbox{}
+\cR \tf \nh
\Bigg[
\frac{704}{9}-\frac{64 \pi ^2}{9}
\Bigg]
    + \ldots
\,,\nonumber\\
  \Delta^{(3)}\Big|_{n_l} &=&
\cR^2 \tf \nl
\Bigg[
\frac{1}{\beta^2} \Bigg(\frac{64}{9} \pi ^4 \log (2 \beta )+\frac{128 \pi ^2
  \zeta (3)}{3}-\frac{160 \pi ^4}{27}\Bigg)+ \frac{1}{\beta}
\Bigg(-\frac{208}{3} \pi ^2 \log (2 \beta )
\nonumber\\&&\mbox{}
+32 \pi ^2 \zeta (3)+\frac{662 \pi ^2}{9}\Bigg)
+ \frac{3584 \afour}{3}-\frac{256}{9} \pi ^2 \log ^2(2 \beta
)+\frac{320}{27} \pi ^4 \log (2 \beta )
\nonumber\\&&\mbox{}
+\frac{7232}{27} \pi ^2 \log (2 \beta )+\frac{640 \pi ^2 \zeta
  (3)}{9}+\frac{19984 \zeta (3)}{9}-\frac{12952 \pi
  ^4}{405}+\frac{8032 \pi ^2}{81}-\frac{416}{9}
\nonumber\\&&\mbox{}
+\frac{448 \logtwo^4}{9}-\frac{128}{3} \pi ^2 \logtwo^2-160 \pi ^2 \logtwo
\Bigg]
\nonumber\\&&\mbox{}
+\cA \cR \tf \nl
\Bigg[
\frac{1}{\beta} \Bigg(-\frac{704}{9} \pi ^2 \log ^2(2 \beta )+\frac{3472}{27}
\pi ^2 \log (2 \beta )-\frac{112 \pi ^2 \zeta (3)}{3}
\nonumber\\&&\mbox{}
-\frac{352 \pi ^4}{27}-\frac{3596 \pi ^2}{81}\Bigg)
-\frac{8960 \afour}{9}-\frac{128}{3} \pi ^2 \log ^2(2 \beta
)+\frac{1552}{9} \pi ^2 \log (2 \beta )
\nonumber\\&&\mbox{}
+\frac{796 \zeta (3)}{9}+\frac{2788 \pi ^4}{81}-\frac{17392 \pi
  ^2}{81}+\frac{78880}{81}-\frac{1120 \logtwo^4}{27}+\frac{1216}{27}
\pi ^2 \logtwo^2
\nonumber\\&&\mbox{}
+\frac{5480}{27} \pi ^2 \logtwo
\Bigg]
+\cR \tf^2 \nl^2
\Bigg[
  \frac{1}{\beta} \Bigg(\frac{128}{9} \pi ^2 \log ^2(2 \beta )
-\frac{640}{27} \pi ^2 \log (2 \beta )
+\frac{64 \pi ^4}{27}
\nonumber\\&&\mbox{}
+\frac{800 \pi ^2}{81}\Bigg)
-\frac{10432}{81}-\frac{512 \pi ^2}{27}
\Bigg]
+ \cR \tf^2 \nh \nl
\Bigg(\frac{3328 \pi ^2}{81}-\frac{35648}{81}\Bigg)
    + \ldots
\,,
\nonumber\\  \label{eq::delta123}
\end{eqnarray}
where the ellipses refer to higher order terms in $\beta$.  
Note that higher order $\epsilon$ terms in the one- and two-loop
expressions are needed to obtain Eq.~(\ref{eq::delta123}) since 
there are products of form factor in Eq.~(\ref{eq::sigee}), which
contain poles in $\epsilon$.
At two
loops the $n_h$ contribution with a closed massive
fermion loop does not develop a $1/\beta$ term since the Coulomb
singularity is regulated by the quark mass in the closed loop. 
For the same reason we
have that the $n_l n_h$ term at three loops starts at ${\cal
  O}(\beta^0)$.  Results for $\Delta^{(3)}$ in the large-$N_c$ limit
can be found in Eq.~(4.9) of Ref.~\cite{Henn:2016tyf}.  The terms
in $\Delta^{(3)}$, which are enhanced by inverse powers of $\beta$,
agree with
Refs.~\cite{Pineda:2006ri,Hoang:2008qy,Kiyo:2009gb}.\footnote{We thank
  Andreas Maier for providing the result for $\Pi^{(3),v}(z)$ in
  Eq.~(A.6) of Ref.~\cite{Kiyo:2009gb} and the corresponding two-loop
  expression in terms of Casimir invariants.}


\subsection{\label{sub::num}Numerical results}

In this subsection we demonstrate the numerical evaluation of our results. We
set $n_l=5$ and consider the $\epsilon^0$ terms for $F_1$ (analogous
results can easily be obtained for $F_2$) and show results for $x\in[-1,1]$
and $\phi\in[0,\pi]$ ($x=e^{i\phi}$) which covers all $q^2$ values on the real
axis.  For $x\in[-1,1]$ we subtract the leading high-energy behaviour, which
contains logarithmic divergences (cf. Eq.~(\ref{eq::fflar})) in order to have
a smooth behaviour for $x\to0$. Thus we define ($i=1,2$)
\begin{eqnarray}
  \hat{F}_i(q^2) = F_i(q^2) - F_i(q^2)\Big|_{q^2\to\infty}
  \,,
\end{eqnarray}
where the second term on the r.h.s. is obtained from the high-energy
expansion discussed in Subsection~\ref{sub::he} by omitting power suppressed
terms. For negative $x$ one should interpret $\log(x)$ as $\log(x+i0)=\log(-x)+i\pi$.

\begin{figure}[t] 
  \begin{center}
    \begin{tabular}{cc}
      \includegraphics[width=0.45\textwidth]{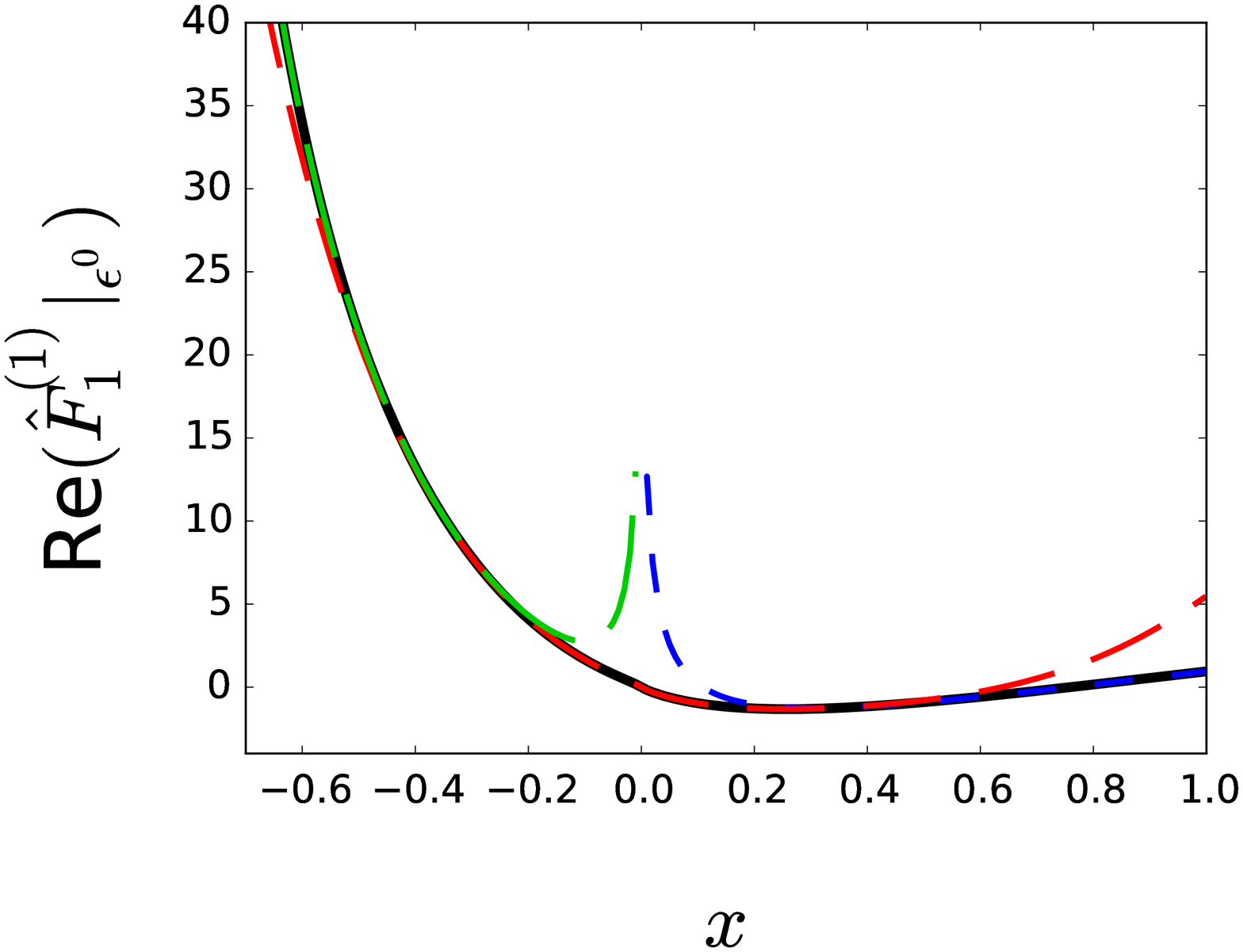} &
      \includegraphics[width=0.45\textwidth]{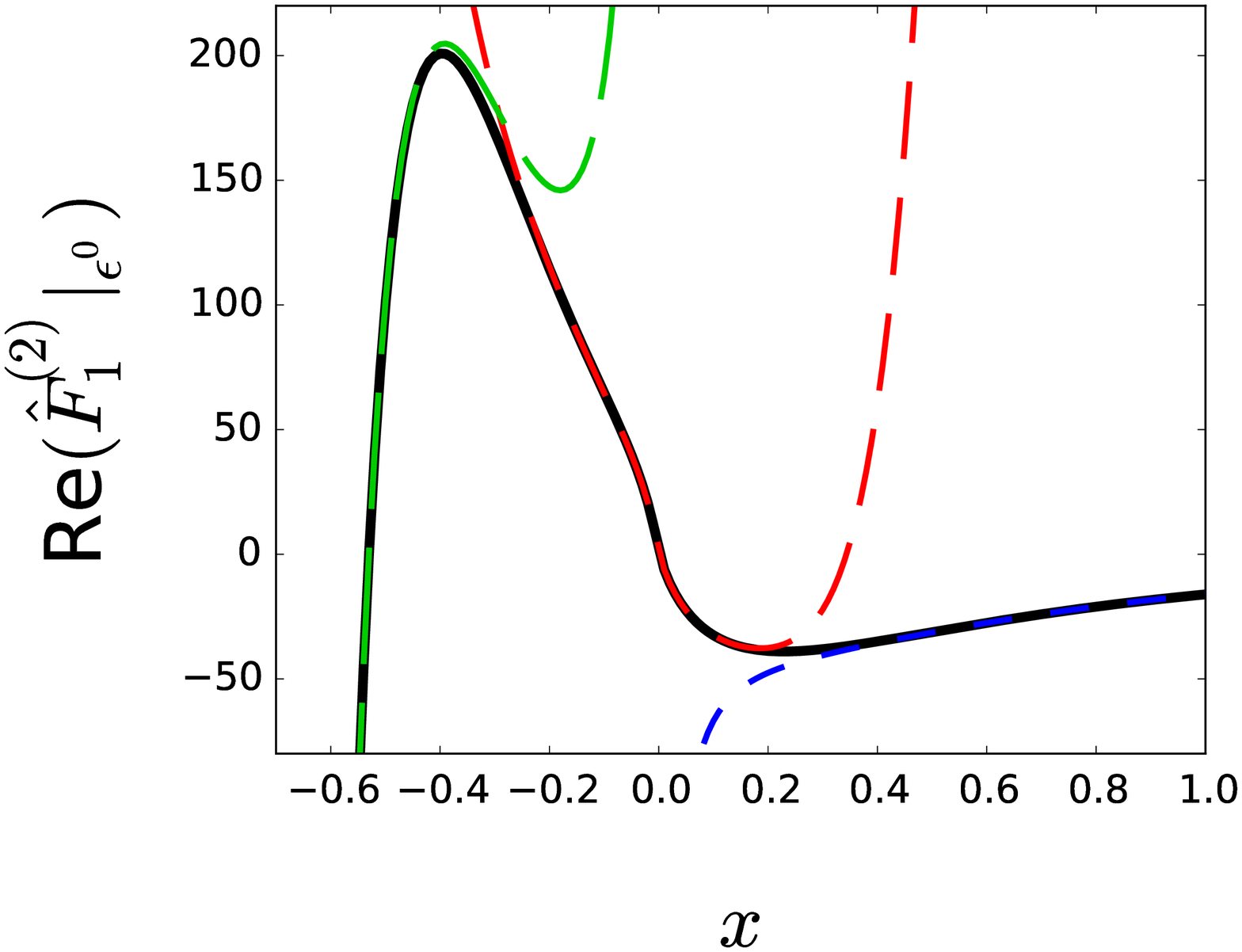} \\
      \\[-1em]
      (a) & (b)
      \\[1em]
      \includegraphics[width=0.45\textwidth]{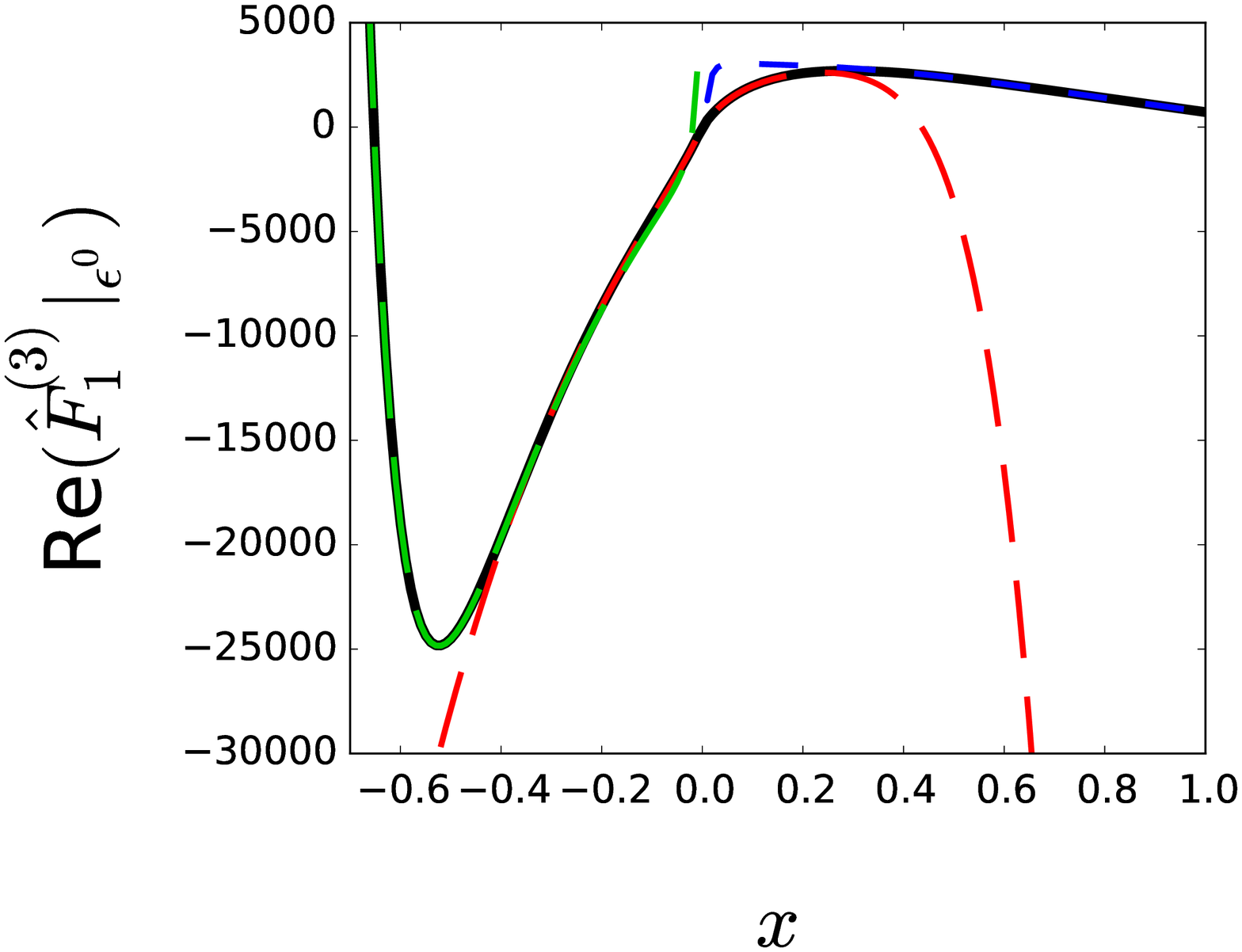}
      \\
      (c)
    \end{tabular}
    \caption{\label{fig::F1num}Panels (a), (b) and (c) show the real
      part of the $\epsilon^0$ term of $\hat{F}_1(q^2)$ at one-, two-
      and three-loop order, respectively. The solid (black) lines
      correspond to the exact result and the dashed lines to
      approximations.}
  \end{center}
\end{figure}

In Fig.~\ref{fig::F1num} the real part of the $\epsilon^0$ term of
$\hat{F}_1(q^2)$ is shown at one, two and three loops.  We show both
the exact result (solid, black curve) and the expansions in the three
kinematic regions (discussed above) as dashed lines.  The approximations
contain terms up to order $x^6$ and $(1-x)^6$ in the high- and
low-energy expansion, respectively. At threshold we include
terms up to order $\beta^4$. The numerical
evaluation of the GPLs is performed with the help of {\tt
  ginac}~\cite{Bauer:2000cp,Vollinga:2004sn}.

In all three cases one observes strong power-like singularities for $x\to-1$
($q^2\to 4 m^2$). For this reason we choose $x=-0.7$ as the lower end of the
$x$-axes. One observes that the threshold expansion (long-dashed, green
curves) reproduces this behaviour and follows the exact curve up to about
$x\approx -0.2, -0.3$ and $-0.1$ at one-, two- and three-loop order,
respectively.  At low energies ($x\to1$) $\hat{F}^{(n)}_1(q^2)$ shows a smooth
behaviour and the corresponding approximation (short-dashed, blue curves)
approximate the exact result up to about $x=0.2$.  Finally, the region around
$x\approx0$ is nicely covered by the high-energy approximation (medium-sized
dashes, red curve) which follows the exact curve from about $x=-0.4$ to
$x=0.2$.  Altogether for each $x$-value in $[-1,1]$ at least one of the
expanded results coincides with the exact curve.

\begin{figure}[t] 
  \begin{center}
    \begin{tabular}{cc}
      \includegraphics[width=0.45\textwidth]{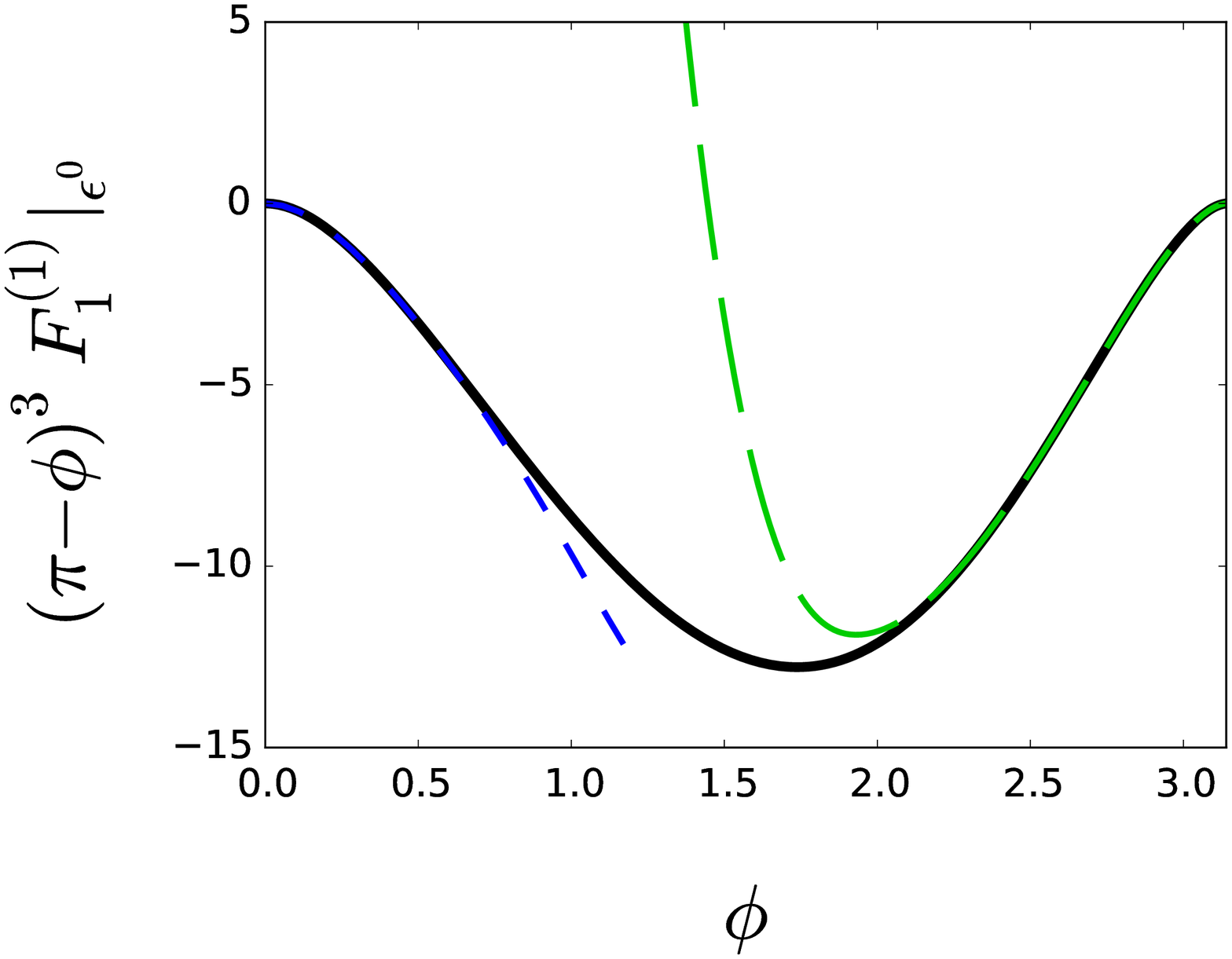} &
      \includegraphics[width=0.45\textwidth]{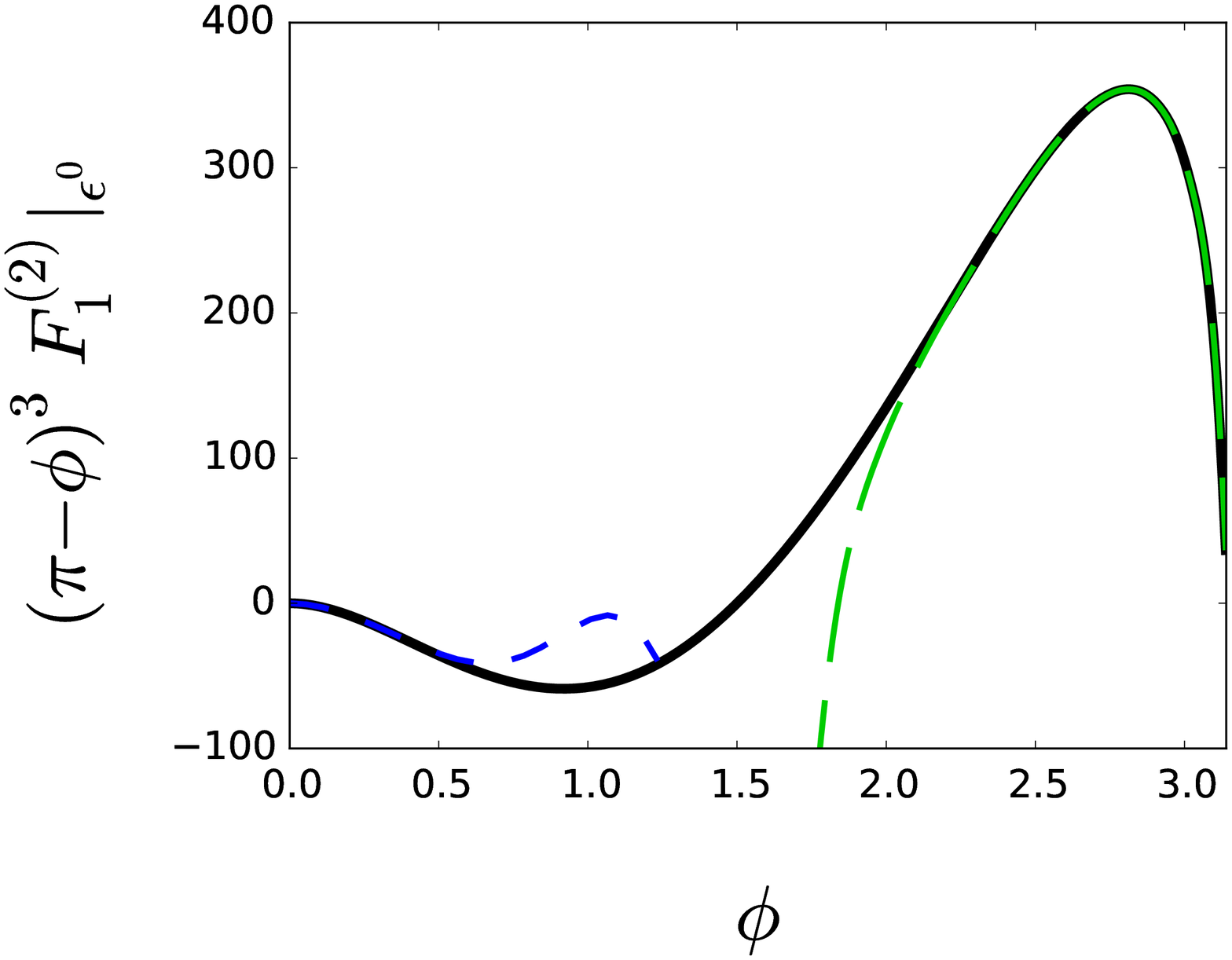} \\
      \\[-1em]
      (a) & (b)
      \\[1em]
      \includegraphics[width=0.45\textwidth]{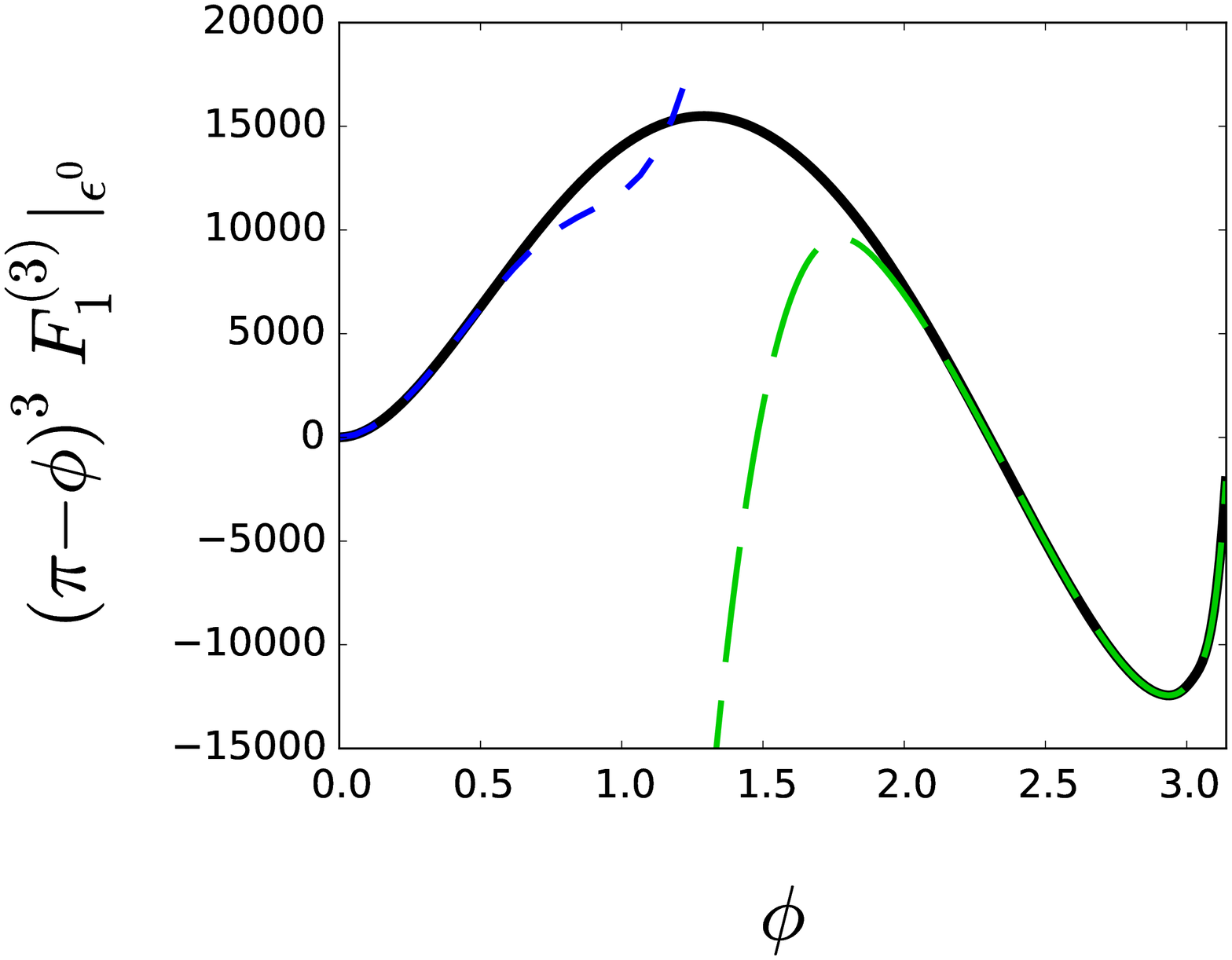}
      \\
      (c)
    \end{tabular}
    \caption{\label{fig::F1num_phi}Panels (a), (b) and (c) show the real
      part of the $\epsilon^0$ term of $(\pi-\phi)^3 F_1(q^2)$ at one-, two-
      and three-loop order, respectively. The solid (black) lines
      correspond to the exact result and the dashed lines to
      approximations.}
  \end{center}
\end{figure}

Fig.~\ref{fig::F1num_phi} shows the real part of the $\epsilon^0$ term of
$(\pi-\phi)^3 F_1(q^2)$ at one-, two- and three-loop order as a function of
$\phi\in[0,\pi]$, which corresponds to the $q^2/m^2$ range between 0 and 4.
We have introduced the factor $(\pi-\phi)^3$ in order to suppress the
singularity at threshold ($\phi\to\pi$). In fact, this factor guarantees that
the one- and two-loop expressions become zero and the $(\pi-\phi)^3 F_1^{(3)}$
is constant for $\phi=\pi$.  As in Fig.~\ref{fig::F1num} we show the exact
result as solid line and the low-energy and threshold approximation as short-
and long-dashed curves. Good agreement is found for 
$\phi\lesssim0.5$ and $\phi\gtrsim2.0$, respectively, which corresponds to
$q^2/m^2\lesssim0.24$ and $q^2/m^2\gtrsim 2.8$. The range inbetween
is not covered by our approximations. In principle we could increase the
expansion depth, which we refrain to do in this paper.

For $\phi\in[0,\pi]$ the form factors $F_1$ and $F_2$ have to be
real-valued. We have verified this feature numerically
which serves as a welcome check for our calculation.
Note that the individual GPLs are complex-valued and the imaginary
parts only cancel in the sum.

We refrain from showing results for the imaginary part of $F_1$ and $F_2$,
which are obtained in a straightforward way using the
expressions in the ancillary file to this paper.
We observe qualitatively similar results as in Figs.~\ref{fig::F1num}
and~\ref{fig::F1num_phi}.

We want to stress that a large part of the $x$ and $\phi$ range is covered by
the approximations in the kinematic regions which have a much simpler structure
than the exact expressions. Thus, if one wants to have a fast numerical
evaluation it is possible to resign to the approximations without essential
loss of precision.



\section{\label{sec::sum}Conclusions and outlook}

In this paper we take the next step in computing massive form factors to
three-loop order and compute the complete light-fermion corrections to the
massive photon quark form factor. We obtain analytic expressions for $F_1$ and
$F_2$.  Our result is expressed in terms of Goncharov polylogarithms with
letters $\pm 1$ and $0$.  This is the first time that non-planar diagrams have
been considered to evaluate massive three-loop vertex functions.

As by-products we compute the two-loop form factors to order $\epsilon^2$ and
confirm the light-fermion part of the three-loop cusp anomalous dimension.

We expand our exact expressions in the low-energy, threshold, and high-energy
limits, and obtain results which are enhanced (for example logarithmically at
high energies or by inverse powers of the velocity at threshold) as well as
power suppressed terms.

The large-$N_c$ results for $F_1$ and $F_2$, which have been computed in
Ref.~\cite{Henn:2016tyf}, are also expressed in terms of Goncharov
polylogarithms, however, an additional fourth letter, $r_1=e^{i\pi/3}$, is
required.  To complete the evaluation of the massive three-loop corrections to
$F_1$ and $F_2$ one has to consider also non-planar non-fermionic
contributions.  It is expected that the corresponding analytic result
leaves the class of GPLs and elliptic integrals appear.  Still, we expect that
fast and flexible numerical evaluations of the form factors are possible,
e.g., with the help of the strategy presented in Ref.~\cite{Lee:2017qql}.


\section*{\label{sec::ack}Acknowledgments}

This work is supported by RFBR, grant 17-02-00175A, and by the Deutsche
Forschungsgemeinschaft through the project ``Infrared and threshold effects in
QCD''.  R.L.  acknowledges support from the ``Basis'' foundation for
theoretical physics and mathematics.  V.S. thanks Claude Duhr for permanent
help in manipulations with Goncharov polylogarithms.  We thank Alexander
Penin for carefully reading the manuscript and for useful comments.



\begin{thebibliography}{99}

%
%

\bibitem{Henn:2016kjz}
  J.~M.~Henn, A.~V.~Smirnov and V.~A.~Smirnov,
  JHEP {\bf 1612} (2016) 144
  [arXiv:1611.06523 [hep-ph]].

\bibitem{Henn:2016tyf}
  J.~Henn, A.~V.~Smirnov, V.~A.~Smirnov and M.~Steinhauser,
  JHEP {\bf 1701} (2017) 074
  [arXiv:1611.07535 [hep-ph]].

\bibitem{Henn:2016men}
  J.~M.~Henn, A.~V.~Smirnov, V.~A.~Smirnov and M.~Steinhauser,
  JHEP {\bf 1605} (2016) 066
  [arXiv:1604.03126 [hep-ph]].

\bibitem{Lee:2016ixa}
  J.~Henn, A.~V.~Smirnov, V.~A.~Smirnov, M.~Steinhauser and R.~N.~Lee,
  JHEP {\bf 1703} (2017) 139
  [arXiv:1612.04389 [hep-ph]].

\bibitem{Lee:2017mip}
  R.~N.~Lee, A.~V.~Smirnov, V.~A.~Smirnov and M.~Steinhauser,
  Phys.\ Rev.\ D {\bf 96} (2017) no.1,  014008
  [arXiv:1705.06862 [hep-ph]].

\bibitem{vonManteuffel:2016xki}
  A.~von Manteuffel and R.~M.~Schabinger,
  Phys.\ Rev.\ D {\bf 95} (2017) no.3,  034030
  [arXiv:1611.00795 [hep-ph]].

\bibitem{Boels:2017skl}
  R.~H.~Boels, T.~Huber and G.~Yang,
  Phys.\ Rev.\ Lett.\  {\bf 119} (2017) no.20,  201601
  [arXiv:1705.03444 [hep-th]].

\bibitem{Boels:2017fng}
  R.~H.~Boels, T.~Huber and G.~Yang,
  arXiv:1712.07563 [hep-th].

\bibitem{Bernreuther:2004ih}
  W.~Bernreuther, R.~Bonciani, T.~Gehrmann, R.~Heinesch, T.~Leineweber,
  P.~Mastrolia and E.~Remiddi,
  Nucl.\ Phys.\ B {\bf 706} (2005) 245
  [hep-ph/0406046].

\bibitem{Gluza:2009yy}
  J.~Gluza, A.~Mitov, S.~Moch and T.~Riemann,
  JHEP {\bf 0907} (2009) 001
  [arXiv:0905.1137 [hep-ph]].

\bibitem{Ahmed:2017gyt}
  T.~Ahmed, J.~M.~Henn and M.~Steinhauser,
  JHEP {\bf 1706} (2017) 125
  [arXiv:1704.07846 [hep-ph]].

\bibitem{Ablinger:2017hst}
  J.~Ablinger, A.~Behring, J.~Blümlein, G.~Falcioni, A.~De Freitas,
  P.~Marquard, N.~Rana and C.~Schneider,
      arXiv:1712.09889 [hep-ph].

\bibitem{Grozin:2017aty}
  A.~Grozin,
  Eur.\ Phys.\ J.\ C {\bf 77} (2017) no.7,  453
  [arXiv:1704.07968 [hep-ph]].

\bibitem{Polyakov:1980ca}
  A.~M.~Polyakov,
  Nucl.\ Phys.\ B {\bf 164} (1980) 171.

\bibitem{Brandt:1981kf}
  R.~A.~Brandt, F.~Neri and M.~a.~Sato,
  Phys.\ Rev.\ D {\bf 24} (1981) 879.

\bibitem{Korchemsky:1987wg}
  G.~P.~Korchemsky and A.~V.~Radyushkin,
  Nucl.\ Phys.\ B {\bf 283} (1987) 342.

\bibitem{Grozin:2014hna}
  A.~Grozin, J.~M.~Henn, G.~P.~Korchemsky and P.~Marquard,
  Phys.\ Rev.\ Lett.\  {\bf 114} (2015) no.6,  062006
  [arXiv:1409.0023 [hep-ph]].

\bibitem{Grozin:2015kna}
  A.~Grozin, J.~M.~Henn, G.~P.~Korchemsky and P.~Marquard,
  JHEP {\bf 1601} (2016) 140
  [arXiv:1510.07803 [hep-ph]].

\bibitem{Smirnov:2014hma}
  A.~V.~Smirnov,
  Comput.\ Phys.\ Commun.\  {\bf 189} (2015) 182
  [arXiv:1408.2372 [hep-ph]].

\bibitem{Lee:2012cn}
  R.~N.~Lee,
  arXiv:1212.2685 [hep-ph].

\bibitem{Lee:2013mka}
  R.~N.~Lee,
  J.\ Phys.\ Conf.\ Ser.\  {\bf 523} (2014) 012059
  [arXiv:1310.1145 [hep-ph]].

\bibitem{Goncharov:1998kja}
  A.~B.~Goncharov,
  Math.\ Res.\ Lett.\  {\bf 5} (1998) 497
  [arXiv:1105.2076 [math.AG]].

\bibitem{Nogueira:1991ex}
  P.~Nogueira,
  J.\ Comput.\ Phys.\  {\bf 105} (1993) 279;\\
  \verb|http://cfif.ist.utl.pt/~paulo/qgraf.html|.

\bibitem{Kuipers:2012rf}
  J.~Kuipers, T.~Ueda, J.~A.~M.~Vermaseren and J.~Vollinga,
  Comput.\ Phys.\ Commun.\  {\bf 184} (2013) 1453
  [arXiv:1203.6543 [cs.SC]].

\bibitem{Harlander:1997zb}
  R.~Harlander, T.~Seidensticker and M.~Steinhauser,
  Phys.\ Lett.\ B {\bf 426} (1998) 125
  [hep-ph/9712228].

\bibitem{Seidensticker:1999bb}
  T.~Seidensticker,
  hep-ph/9905298.

\bibitem{Smirnov:2013dia}
  A.~V.~Smirnov and V.~A.~Smirnov,
  Comput.\ Phys.\ Commun.\  {\bf 184} (2013) 2820
  [arXiv:1302.5885 [hep-ph]].

\bibitem{Lee:2014ioa}
  R.~N.~Lee,
  JHEP {\bf 1504} (2015) 108
  [arXiv:1411.0911 [hep-ph]].

\bibitem{Remiddi:1999ew}
  E.~Remiddi and J.~A.~M.~Vermaseren,
  Int.\ J.\ Mod.\ Phys.\ A {\bf 15} (2000) 725
  [hep-ph/9905237].

\bibitem{Lee:2010ik}
  R.~N.~Lee and V.~A.~Smirnov,
  JHEP {\bf 1102} (2011) 102
  [arXiv:1010.1334 [hep-ph]].

\bibitem{progdata}
\verb|https://www.ttp.kit.edu/preprints/2018/ttp18-006/|.

\bibitem{Grozin:2007fh}
  A.~G.~Grozin, P.~Marquard, J.~H.~Piclum and M.~Steinhauser,
  Nucl.\ Phys.\ B {\bf 789} (2008) 277
  [arXiv:0707.1388 [hep-ph]].

\bibitem{Sudakov:1954sw}
  V.~V.~Sudakov,
  Sov.\ Phys.\ JETP {\bf 3} (1956) 65
   [Zh.\ Eksp.\ Teor.\ Fiz.\  {\bf 30} (1956) 87].

\bibitem{Frenkel:1976bj}
  J.~Frenkel and J.~C.~Taylor,
  Nucl.\ Phys.\ B {\bf 116} (1976) 185.

\bibitem{Penin:2014msa}
  A.~A.~Penin,
  Phys.\ Lett.\ B {\bf 745} (2015) 69
   Erratum: [Phys.\ Lett.\ B {\bf 751} (2015) 596]
   Erratum: [Phys.\ Lett.\ B {\bf 771} (2017) 633]
   10.1016/j.physletb.2015.10.035
  [arXiv:1412.0671 [hep-ph]].

\bibitem{Liu:2017axv}
  T.~Liu, A.~A.~Penin and N.~Zerf,
  Phys.\ Lett.\ B {\bf 771} (2017) 492
  [arXiv:1705.07910 [hep-ph]].

\bibitem{Liu:2017vkm}
  T.~Liu and A.~A.~Penin,
  Phys.\ Rev.\ Lett.\  {\bf 119} (2017) no.26,  262001
  [arXiv:1709.01092 [hep-ph]].

\bibitem{Pineda:2006ri}
  A.~Pineda and A.~Signer,
  Nucl.\ Phys.\ B {\bf 762} (2007) 67
  [hep-ph/0607239].

\bibitem{Hoang:2008qy}
  A.~H.~Hoang, V.~Mateu and S.~Mohammad Zebarjad,
  Nucl.\ Phys.\ B {\bf 813} (2009) 349
  [arXiv:0807.4173 [hep-ph]].

\bibitem{Kiyo:2009gb}
  Y.~Kiyo, A.~Maier, P.~Maierh\"ofer and P.~Marquard,
  Nucl.\ Phys.\ B {\bf 823} (2009) 269
  [arXiv:0907.2120 [hep-ph]].

\bibitem{Bauer:2000cp}
  C.~W.~Bauer, A.~Frink and R.~Kreckel,
  J.\ Symb.\ Comput.\  {\bf 33} (2000) 1
  [cs/0004015 [cs-sc]].

\bibitem{Vollinga:2004sn}
  J.~Vollinga and S.~Weinzierl,
  Comput.\ Phys.\ Commun.\  {\bf 167} (2005) 177
  [hep-ph/0410259].

\bibitem{Lee:2017qql}
  R.~N.~Lee, A.~V.~Smirnov and V.~A.~Smirnov,
  arXiv:1709.07525 [hep-ph].


\end{thebibliography}
\end{document}